\documentclass[10pt,conference]{IEEEtran}

\usepackage{url}

\usepackage{listings}
\usepackage{bm}

\usepackage{xspace}
\usepackage{amsmath} 
\usepackage{soul}

\usepackage{verbatim}
\usepackage{tcolorbox}

\usepackage{stfloats}
\usepackage{textcomp}
\usepackage{booktabs, tabularx}
\usepackage{rotating}
\usepackage{xcolor}
\usepackage{color,colortbl}
\usepackage{soul}

\usepackage{fancybox}
\usepackage{multirow}
\usepackage{multicol}

\usepackage{url}

\usepackage[caption=false,font=footnotesize]{subfig}
\usepackage{balance}
\usepackage{footnote}

\newcommand\noindexfootnote[1]{%
  \begingroup
  \renewcommand\thefootnote{}\footnote{#1}%
  \addtocounter{footnote}{-1}%
  \endgroup
}

\usepackage{array}

\usepackage{adjustbox}
\usepackage{threeparttable}

\definecolor{dkgreen}{rgb}{0,0.6,0}
\definecolor{gray}{rgb}{0.5,0.5,0.5}
\definecolor{mauve}{rgb}{0.58,0,0.82}
\definecolor{dkred}{rgb}{0.6, 0.1, 0}
\definecolor{dkblue}{rgb}{0, 0, 0.6}

\newcommand{\tool}{NLPerturbator\xspace}

\newcommand{\phead}[1]{\vspace{1mm} \noindent {\bf #1}}
\newcommand{\stage}[1]{\textbf{\em #1}}
\newcommand{\exampleo}[1]{\textcolor{blue}{\textbf{#1}}}
\newcommand{\examplep}[1]{\textcolor{red}{\textbf{#1}}}
\newcommand{\down}[1]{\textcolor{blue}{\textit{#1}}}
\newcommand{\sdown}[1]{\textcolor{blue}{\textit{\textbf{#1}}}}

\newcommand{\rqbox}[1]{\begin{tcolorbox}[left=2pt,right=2pt,top=2pt,bottom=2pt,colback=gray!5,colframe=gray!40!black,before skip=2pt,after skip=2pt]#1\end{tcolorbox}}

\usepackage{array}

\usepackage[normalem]{ulem}
\usepackage{pifont}
\usepackage{enumitem}

\definecolor{lred}{RGB}{255,193,192}
\definecolor{lgreen}{RGB}{204,255,216}

\begin{document}

\title{NLPerturbator: Studying the Robustness of Code LLMs to Natural Language Variations}

\makeatletter
\newcommand{\linebreakand}{%
  \end{@IEEEauthorhalign}
  \hfill\mbox{}\par
  \mbox{}\hfill\begin{@IEEEauthorhalign}
}
\makeatother

\author{
  \IEEEauthorblockN{Junkai Chen*}
  \IEEEauthorblockA{
    Zhejiang University\\
    Ningbo, China \\
    junkaichen@zju.edu.cn}
  \and
  \IEEEauthorblockN{Zhenhao Li*}
  \IEEEauthorblockA{
    York University\\
    Toronto, Canada \\
    zhenhao.li@ieee.org}
  \and
  \IEEEauthorblockN{Xing Hu$^{\dagger}$}
  \IEEEauthorblockA{
    Zhejiang University\\
    Ningbo, China \\
    xinghu@zju.edu.cn}

  \and
  \IEEEauthorblockN{Xin Xia}
  \IEEEauthorblockA{
    Zhejiang University\\
    Hangzhou, China \\
    xin.xia@acm.org}

}

\maketitle

\begin{abstract}
Large language models (LLMs) achieve promising results in code generation based on a given natural language description. They have been integrated into open-source projects and commercial products to facilitate daily coding activities.
The natural language description in the prompt is crucial for LLMs to comprehend users' requirements.
Prior studies uncover that LLMs are sensitive to the changes in the prompts, including slight changes that look inconspicuous. However, the natural language descriptions often vary in real-world scenarios (e.g., different formats, grammar, and wording). Prior studies on the robustness of LLMs are often based on random perturbations and such perturbations may not actually happen. In this paper, we conduct a comprehensive study to investigate how are code LLMs robust to variations of natural language description in real-world scenarios. We summarize 18 categories of perturbations of natural language and 3 combinations of co-occurred categories based on our literature review and an online survey with practitioners. We propose an automated framework, \tool, which can perform perturbations of each category given a set of prompts. Through a series of experiments on code generation using six code LLMs, we find that the perturbed prompts can decrease the performance of code generation by a considerable margin (e.g., up to 21.2\%, and 4.8\% to 6.1\% on average). Our study highlights the importance of enhancing the robustness of LLMs to real-world variations in the prompts, as well as the essentiality of attentively constructing the prompts.

\noindexfootnote{* Co-first author, equally contributed.}
\noindexfootnote{$^{\dagger}$ Corresponding author.}

\end{abstract}


\vspace{-0.5cm}
\section{Introduction}
\label{sec:intro}

Large language models (LLMs) have shown impressive and promising capabilities in software engineering tasks~\cite{liu2023fill,feng2023prompting,geng2023large}. Among such tasks, code generation is an important practice in AI-assisted software engineering and has been widely studied by prior LLM-related works~\cite{codex,starcoder,llama2,lin2024llm,icse24_code_suggestion_rag}. Given the prompt with natural language description, LLMs generate the source code correspondingly. Practitioners can then utilize the generated code to accelerate the process of development and maintenance.

The natural language (NL) description in the prompt is crucial to convey the requirements defined by users to LLMs. Prior studies evaluate the capability of LLMs on code generation based on the datasets with human-verified prompts. However, the NL description in the prompts may vary in real-world scenarios due to different wording, grammar, format, and even typos. Prior studies find that LLMs are usually sensitive to such variations~\cite{shen2023chatgpt, wang2023robustness}, a slight change may lead to a completely different result. For example, Figure ~\ref{fig:motivation} shows a pair of code snippets generated by StarCoder~\cite{starcoder}. Using the original prompt in HumanEval~\cite{codex} dataset, StarCoder generates a correct solution code snippet. However, StarCoder generates an incorrect code snippet by simply repeating a word {\em ``it''} in the prompt. Such variations or perturbations of NL descriptions may often occur in real-world scenarios.

Prior studies proposed a series of approaches to evaluate the robustness of LLMs~\cite{wang2022adversarial, zhu2023promptbench}. However, such approaches often focus on natural language processing (NLP) tasks (e.g., sentiment analysis) and are mostly based on random perturbations (e.g., randomly replacing characters in random locations). Therefore, it is still unclear how are LLMs robust to real-world variations in NL descriptions when practitioners are using LLMs to perform code generations.

In this paper, we conduct a comprehensive study to investigate how are code LLMs robust to perturbations of NL description in real-world scenarios: 1) we first derive a list of 25 initial categories of perturbations in NL based on prior studies and our experience in using LLMs to generate code; 2) we conduct an online survey with practitioners in the open-source community, industry, and academia to verify if the initial categories of perturbations can actually occur in real-world scenarios. We have a list of 18 final categories and several combinations of co-occurred categories as the result; 3) we propose an automated framework, \tool, which can parse the NL description from the prompt and then apply perturbations of specific categories including the 18 final categories and 3 combinations of co-occurred categories; 4) we conduct experiments on six open-source code LLMs to evaluate their robustness to the perturbed prompts.

\begin{figure}
 \centering
\includegraphics[width=1.00\linewidth]{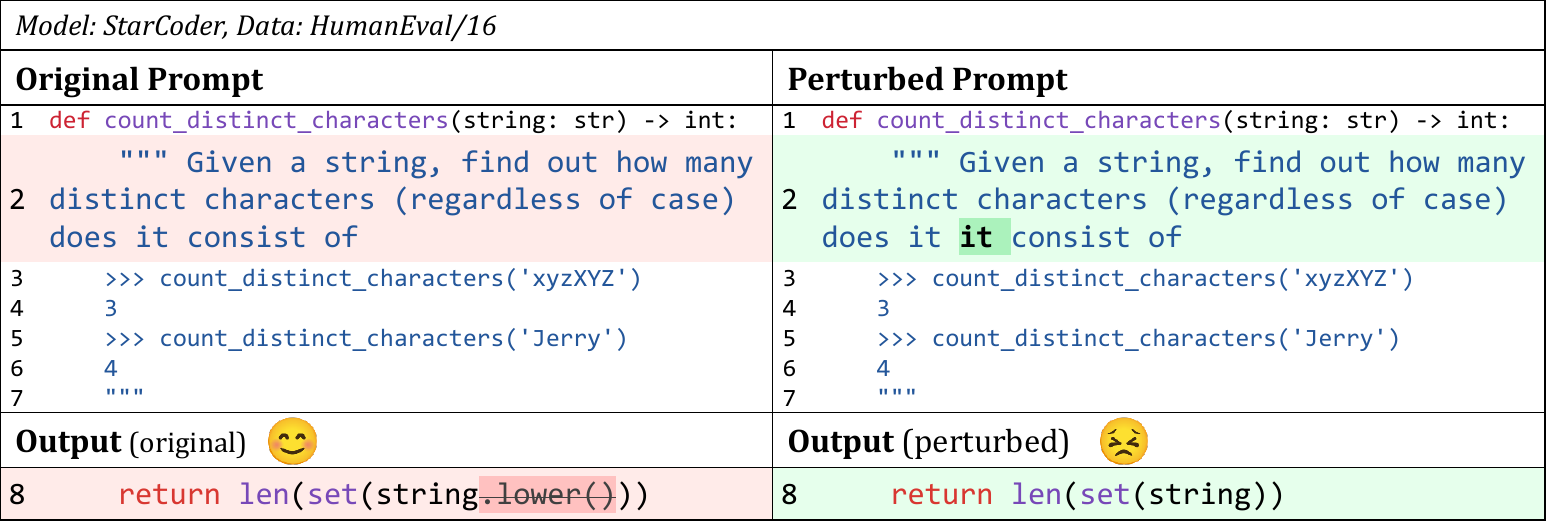}
 \caption{Example of code generation using original and perturbed prompts. StarCoder~\cite{starcoder} fails to generate a correct solution (i.e., neglects the requirement of letter case) when an extra word ``it'' is repeated in the natural language description of the problem \textit{HumanEval/16}.}.
 \label{fig:motivation}
 \vspace{-0.1cm}
 \end{figure}

In particular, we study the following three research questions:

\phead{RQ1: What are the Practitioners' Perspective on the Categories of Perturbations in Natural Languages?}
In this research question, we analyze the survey results for their perspective on each initial category of perturbation and their feedback on co-occurred categories. We find that 18 out of the 25 initial categories have a majority of survey participants considering that they can occur in real-world scenarios. Moreover, 86.5\% of the participants consider that multiple categories of perturbations can occur in the same prompt. They provide 80 unique combinations of co-occurred categories in total.

\phead{RQ2: How are the Results of Automatically Composing Perturbations in Natural Languages that Reflect Real-world Scenarios?}
In this research question, we manually examine the prompts perturbed by \tool for each category. Specifically, we examine the effectiveness (i.e., if the perturbation belongs to the desired category and is applied to the NL part of the prompt) and naturalness (i.e., if the resulted prompt looks natural) of each perturbed prompt. We find that \tool has an average effectiveness and naturalness of 94.0\% \& 86.6\% for HumanEval and 96.7\% \& 91.8\% for MBPP, respectively.

\phead{RQ3: How are Code LLMs Robust to the Perturbations in Natural Languages on Code Generation?}
We investigate the robustness of code LLMs to NL perturbations using our manually verified perturbations datasets.
We find that our perturbed prompts can decrease the performance of code generation, while the extent varies for different categories and code LLMs.

We summarize the contributions of this paper as follows:
\begin{itemize}[leftmargin=*]
    \item We propose an automated framework, \tool, which can perform perturbations that reflect real-world scenarios on the NL description of prompts based on the feedback from practitioners.
    \item We provide manually verified datasets, {\sf HumanEval-R} and {\sf MBPP-R}, which can be leveraged to investigate the robustness of LLMs to NL perturbations on code generation.
    \item We conduct a comprehensive evaluation of the robustness of six code LLMs to the NL perturbations and discuss the implications to prompt engineering and directions for improving the robustness of LLMs.

\end{itemize}

\phead{Paper Organization.}
 Section~\ref{sec:related} summarizes the related work. 
 Section~\ref{sec:methodology} discusses the research methodology. 
 Section~\ref{sec:categories} describes the categories of perturbations and the implementation of our perturbators.
 Section~\ref{sec:results} presents the results by answering three research questions.
 Section~\ref{sec:discussion} discusses the implications of our study and the threats to validity. 
 Section~\ref{sec:conclusion} concludes the paper.

\vspace{-0.1cm}

\vspace{-0.1cm}
\section{Related Work}
\label{sec:related}

\phead{Studies on Robustness of LLMs.}
Large language models (e.g., ChatGPT~\cite{chatgpt}) have received extensive attention for their superior understanding and generative abilities. 
Given the importance of robustness for LLMs, a series of researches have been conducted to study such aspect.
Wang et al.~\cite{wang2022adversarial} and Zhu et al.~\cite{zhu2023promptbench} presented benchmarks to evaluate the robustness of LLMs. They compiled a collection of perturbation approaches (e.g., back translation), natural language processing tasks (NLP tasks, e.g., NL translation), datasets (e.g., GLUE), and evaluation metrics (e.g., performance drop rate) to evaluate the robustness of language models. 
Some prior studies discussed the reliability and potential risks of LLMs, including the robustness to various prompts~\cite{shen2023chatgpt,ye2023assessing}.
Apart from NLP tasks, some prior studies~\cite{wang2023recode, mastropaolo2023robustness,zhong2023study,zhou2022adversarial,zhuo2023robustness,shirafuji2023exploring} also explored the robustness of language models in coding-related tasks (e.g., code generation, code completion, and code comment generation). These works followed perturbation approaches and evaluation methodologies from existing NLP studies and discussed the robustness of code models. 
However, such prior studies leveraged perturbation approaches that are almost arbitrary and random, which can hardly reflect real-world scenarios when using LLMs in code tasks.

Different from prior studies, we focus on natural language perturbations that can happen in real-world scenarios when using LLMs to generate code, which may provide more practical insights to comprehend the robustness of LLMs.

\phead{Studies on Code Generation.} 
Previous works proposed many approaches to assist in code generation. We summarize them into approaches based on deep neural networks~\cite{ling2016latent, yin2017syntactic, svyatkovskiy2020intellicode}, pre-trained language models~\cite{svyatkovskiy2020intellicode, wang2021codet5, le2022coderl}, and LLMs~\cite{starcoder, codegen2, codegeex, codellama}.
(1) Deep neural networks have been applied to code generation, which is a preliminary step towards automated code generation. 
For example, Ling et al.~\cite{ling2016latent} and Sun et al.~\cite{sun2020treegen} proposed different neural architectures for code generation that leveraged structural information extracted from the inputs and achieved promising results. 
(2) Pre-trained language models acquire knowledge from large-scale datasets and improve the performance of code generation by a considerable margin. 
Svyatkovskiy et al.~\cite{svyatkovskiy2020intellicode} and Wang et al.~\cite{wang2021codet5} used pre-trained language models to generate code and evaluated the textual similarity (e.g., BLEU score~\cite{bleu}) with the ground truth.
(3) Recently, utilizing LLMs to generate code using natural language descriptions has attracted interest from both industry and academia. 
A variety of code LLMs have shown great effectiveness in code generation, including open-source LLMs (e.g., StarCoder~\cite{starcoder} and CodeGen2~\cite{codegen2}) and commercial LLMs (e.g., Codex~\cite{codex}). These code LLMs vary in many aspects such as model size, training method, and data collection. 
In addition, some works aimed to enhance code LLMs with different techniques such as retrieval~\cite{zhou2022docprompting} and in-context learning~\cite{li2023towards}.

On top of such code generation techniques, we study the natural language perturbations that happen in real-world scenarios. The purpose of our study is to help strengthen the robustness of code LLMs in practice.

\vspace{-0.2cm}
\section{Methodology}
\label{sec:methodology}
Figure~\ref{fig:overall} presents an overview of the methodology of our study. 
\stage{Stage 1:} We derive an initial list of categories of the perturbations in natural language (NL) that might be related to the description of code generation.
\stage{Stage 2:} We conduct an online survey with practitioners for their perspectives on the perturbations that may actually happen in real-world scenarios for code generation.
\stage{Stage 3:} Based on the feedback of our survey, we build a framework that can automatically compose NL perturbations on prompts that are close to real-world scenarios.
\stage{Stage 4:} We investigate how robust code LLMs are to the perturbations on prompts by comparing the results of code generation using original and perturbed prompts.

\begin{figure*}
 \centering
\includegraphics[width=1.00\linewidth]{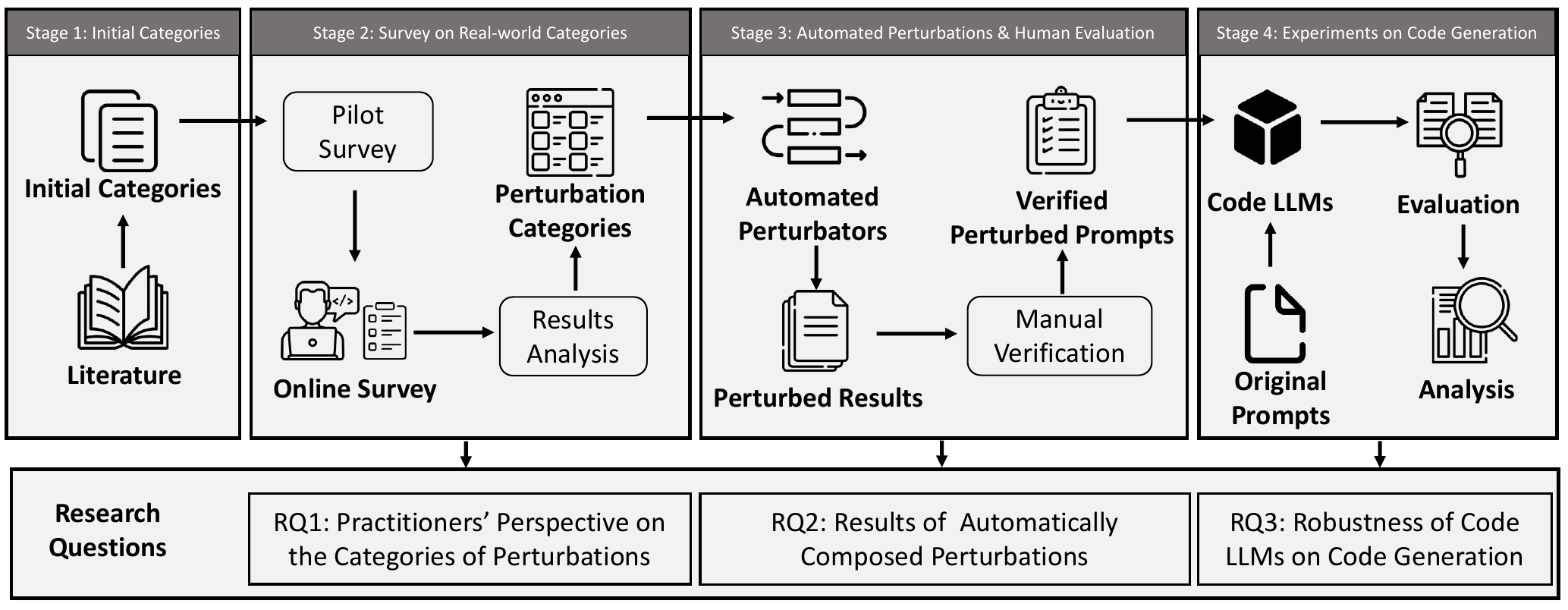}
\vspace{-0.2cm}
 \caption{Overview of our study.}
 \label{fig:overall}
 \vspace{-0.3cm}
 \end{figure*}

\vspace{-0.1cm}
\subsection{Stage 1: Initial Categories of Perturbations in Natural Language}

We focus on studying the NL description of a prompt. In this stage, we manually derive a list of initial categories of perturbations in NL based on prior studies and our experience in using LLMs to generate code.

\phead{Paper Collection.}
To collect a set of papers related to the robustness of LLMs in code generation, we first use the keywords ``robustness code generation'', ``robust code generation'', ``robustness large language models'', ``robust large language models'', ``robustness LLM'',  and ``robust LLM'' to search in Google Scholar. We perform such search strategy since LLM-related work is an emerging topic, many related studies are released on arXiv~\cite{arxiv} as pre-prints before the acceptance from a research venue.
We read the title and abstract of the top 20 search results by relevance and then collect papers related to our study. All of the collected papers appear in the top-20 search results, we do not find the results relevant from top-21 to top-30 search results. We then check the references of these papers to identify more related papers. Eventually, we collect 43 papers that are related to the robustness of language models.  For the collection strategy of these papers, 35 of them are collected from the search results (31 papers are collected from the top-10 search results, 4 papers are collected from the top-11 to top-20 search results) and 8 of them are collected from their references. For the research venues, 15 of the collected papers were published at a research venue (i.e., a conference or a journal) and 28 of them are ArXiv papers (as of October 2023). A full list of the collected papers can be found in our replication package~\cite{replication_package}.

\phead{Derivation of Initial Categories.} Two authors of this paper first independently summarize a list of categories of perturbations based on the collected papers and their experience in using LLMs to generate code. The two authors then combine the independently derived categories together. 
For each category in the combined list, the two authors further independently label whether its definition is clear and can be related to the NL description of code generation or not. 
Following prior studies~\cite{ding2023temporal,deeplv,wheretolog_ASE20}, we use Cohen's Kappa~\cite{kappa} to measure the agreement of the label results between the two authors. Overall, we have a Cohen's Kappa value of 0.63, which indicates a substantial agreement. Subsequently, the two authors discuss the disagreements to reach a consensus, merge similar categories, and derive the list of 25 initial categories of perturbations in NL for code generation as the result.

\subsection{Stage 2: Survey on Perturbations in Real-world Scenarios of Code Generation}

We derive initial categories of perturbations in {\em Stage 1}. To further investigate what kind of perturbations can happen in real-world scenarios for code generation, we conduct an anonymous, online survey with practitioners.
We follow Kitchenham and Pfleeger's guidelines~\cite{survey_guide_2008} and other survey-related studies in Software Engineering~\cite{hu2022practitioners,naming_methods} to conduct the survey.

\subsubsection{\textbf{Survey Design and Protocol.}}
The survey includes three parts and we briefly discuss the design of questions for each part. The complete design of our survey can be found in our replication package~\cite{replication_package}.

\phead{Part 1: Information of Participants.} We ask the participants for their basic information, including the country or region of residence, current occupation, primary job role, and years of experience in the primary job role.  We also ask the participants whether they have experience in using LLMs to generate code.

\phead{Part 2: Naturalness of Categories of Perturbations.} For each category of perturbations that we derive in {\em Stage 1}, we provide an example and ask the participants to rate if the category is natural and may occur in real-world scenarios of code generation using LLMs. The participant can choose from ``Yes'', ``No'', and ``Not sure''. We also provide additional fields of comments where the participants are free to leave their comments related to the categories.

\phead{Part 3: Co-occurrence of Different Categories of Perturbations.} We first ask the participants for whether they consider that multiple categories of perturbations can occur in the same prompt, they can choose from ``Yes'', ``No'', and ``Not sure''. 

\begin{itemize}[leftmargin=*]
    \item {\em If the answer is ``Yes''}: Another short-answer question will pop up to ask for the possible combinations. We provide an instruction for how to input the combinations. We suggest the participants list several combinations (e.g., 2 - 5) that may possibly happen in real-world scenarios and do not need to exhaustively list all the possible combinations. This short-answer question is optional, the participants can leave it blank if they consider the co-occurrence of different categories possible but don't have a clear idea of the specific combinations at the moment.
    \item {\em If the answer is ``No''}: Another short-answer question will pop up to ask for the potential reasons. This is an optional question.
    \item {\em If the answer is ``Not sure''}: No follow-up questions.

\end{itemize}

After the completion of these three parts, we also provide an optional comment field where the participants can leave their additional comments regarding the survey.

\phead{Survey Implementation.} We use Microsoft Forms~\cite{microsoft_forms} to implement the survey following the design discussed above. The survey is anonymous and can be accessed using the link. We conduct a pilot survey with five practitioners to collect their feedback on the overall design of our survey. All the practitioners in the pilot survey have experience in using LLMs to generate code. Their feedback is related to two aspects: (1) clarity and consistency of some terms; and (2) conciseness of the instructions. We make minor modifications to adjust the terms and refine the instructions based on the feedback and then have a final version of the survey. 
The final version of our survey can be found in our replication package~\cite{replication_package}.

\subsubsection{\textbf{Recruitment of Participants.}}
We contact professionals in industry or research institutions and ask for their help to disseminate our survey. Specifically, we send emails to our contacts in Microsoft, Ericsson, Morgan Stanley, Huawei, and other research institutions, encouraging them to disseminate our surveys to their colleagues.

\phead{Demographics.}
In total, we receive 57 survey responses. Our survey participants reside in nine countries or regions across three continents, including 34 participants residing in Asia, 4 participants in Europe, and 19 participants in North America. Their years of experience vary from 1 year to 10 years, with an average of 3.6 years. The top two occupations of the participants are industrial/freelance professional (40\%) and academic/industrial researcher (33\%).

\subsubsection{\textbf{Data Analysis.}}

Here we analyze the data of our survey:
\begin{itemize}[leftmargin=*]
    \item For Part 1, we specifically examine how many participants have experience in using LLMs to generate code. Among the 57 responses, we drop 5 responses that select no experience in using LLMs to generate code. We consider the remaining 52 responses as valid responses for further analysis.
    \item For each category in Part 2, we count the percentage of each answer for if it is natural and can happen in real-world scenarios of code generation using LLMs. The category with a majority of answers as ``Yes'' will be considered into the final list of categories and studied in the remaining stages. 
    \item For Part 3, we count the frequency of each combination of co-occurrences mentioned by the survey participants and select the top 3 combinations for the remaining stages of our study.
\end{itemize}

After this stage, we have a final list of categories that take the survey feedback into account on top of the initial categories.

\vspace{-0.1cm}
\subsection{Stage 3: Automatically Composing Natural Language Perturbations on Prompts}
We automatically compose NL perturbations on prompts based on the categories derived from the previous two stages. 
We further conduct a human evaluation to verify these prompts.

\phead{Datasets.} We conduct the study on two datasets: HumanEval~\cite{codex} and MBPP~\cite{mbpp_austin2021program}, which are widely used by prior studies to evaluate the performance of code LLMs~\cite{starcoder,codellama,codegen2,codegeex,chen2024evaluating}. Table~\ref{tab:datasets} shows the statistics of the datasets.
\begin{itemize}[leftmargin=*]
    \item \textbf{HumanEval}~\cite{codex} is a set of 164 hand-written python programming problems. Each problem includes an input prompt, a canonical solution, and a series of test cases. The input prompt is composed of a function signature, a functional description, and several example outputs.
    \item \textbf{MBPP}~\cite{mbpp_austin2021program} contains 974 coding problems in Python, including a subset of 427 problems that are manually verified by the authors. Each problem includes an input prompt containing the functional description, solution code for reference, and test cases. We leverage the manually verified subset to conduct our study.
\end{itemize}

 \begin{table}
 \tabcolsep=10pt
 \centering
 \caption{Statistics of the datasets.}
  \vspace{-0.2cm}
 \resizebox{\columnwidth}{!} {
 \begin{tabular}{l|c|c} 
 \hline
 \rowcolor[rgb]{0.937,0.937,0.937} Dataset & \textbf{HumanEval} & \textbf{MBPP}  \\ 
 \hline
 Num. of examples                          & 164                & 427            \\
 Avg. characters in NL descriptions         & 231.6             & 88.2          \\
 Avg. words in NL descriptions         & 47.0             & 17.9          \\
 \hline
 \end{tabular}

  }
 \label{tab:datasets}
  \vspace{-0.4cm}
 \end{table}

\phead{Automated Perturbations for each Category.}
We propose an automated framework, \tool, to parse the NL description from the prompt, and then apply perturbations of specific categories (i.e., perturbators) on the NL description in the prompt. Our framework can be configured through the following aspects:

\begin{itemize}[leftmargin=*]
    \item \textbf{Perturbator.} The perturbations that can be applied to the prompts include each of the survey-verified 18 categories as well as each of the top 3 combinations of co-occurred categories. 
    Each perturbator implements the perturbations of a specific category.
    \item \textbf{Times.} Times of perturbations to be applied for each prompt, which can be computed based on the product of a base frequency and the number of perturbable locations. The minimum value is 1 to prevent void perturbations.
    \item \textbf{Dataset.} The dataset to use our framework (e.g., HumanEval).   
\end{itemize}

Given the dataset, our framework automatically identifies the NL part in the prompt, applies the specified perturbators for specific times, and then returns the perturbed dataset. 
We receive suggestions from the survey participants in {\em Stage 2} regarding the implementation of some categories. We take account of their comments while implementing the perturbators.
We will discuss the implementation details of each category in Section~\ref{sec:categories}.

\phead{Human Evaluation on the Perturbed Prompts.}
We conduct a human evaluation to examine the automated perturbed results of our framework. Specifically, we use our framework to generate perturbed results for each category on the two datasets. We set a default frequency of perturbation for each category in both standalone (i.e., each category occurs once in each prompt) and co-occurred (i.e., each category in the combination occurs once in each prompt) scenarios. 
Following a similar process of prior studies on manually annotating the data~\cite{huo2023autolog,li2023ase,li2023icse}, 
two authors of this paper then independently and manually inspect all the results. For each perturbed prompt, the authors examine and label its effectiveness and naturalness:
\begin{itemize}[leftmargin=*]
    \item \textbf{Effectiveness.} If the perturbation belongs to the desired category and is applied to the natural language part of the prompt.
    \item \textbf{Naturalness.} If the resulting prompt looks natural and can occur in real-world scenarios.
\end{itemize}

When the labeling is finished, the two authors then compare their results and discuss each disagreement until they reach a consensus~\cite{dlfinder_TSE}. We have a Cohen's Kappa~\cite{kappa} value of 0.74 in this process, which indicates a substantial agreement.
For the perturbed prompt that is labeled as either not effective or not natural, we manually modify the perturbed prompt to be effective and natural.  

After this stage, we have two datasets with manually-verified perturbations of each category based on HumanEval and MBPP to examine the robustness of LLMs on code generation. We name the two manually-verified datasets as {\sf HumanEval-R} and {\sf MBPP-R}. We will discuss the results of human evaluation in RQ2.

\subsection{Stage 4: Experiments on Code Generation}
We conduct experiments to compare the results of code LLMs on code generation, using the original prompts without perturbations and our perturbed prompts. We leverage {\sf HumanEval-R} and {\sf MBPP-R} derived from the previous stage as the perturbed prompts. Specifically, we feed the prompts with perturbations in each category into the code LLMs and compare their Pass@1 with the results using original prompts.

\phead{Code LLMs.} We use six open-source code LLMs to conduct the experiments on code generation. 
\begin{itemize}[leftmargin=*]
    \item \textbf{InCoder}~\cite{fried2022incoder} is a code LLM that supports both program synthesis and editing. We use the 6B version of Incoder in our study. 
    \item \textbf{Code Llama}~\cite{codellama} is a series of code LLMs further trained based on Llama 2~\cite{llama2} using code tokens. The code Llama family models include several variants (e.g., standard, python, and instruct) and different sizes (e.g., 7B, 13B, 34B). We use the standard model with a 7B size of parameters in our study.
    \item \textbf{CodeGeeX2}~\cite{codegeex} is a multilingual code LLM that supports prompts written in both English and Chinese. It has 6B parameters. 
    \item \textbf{CodeGen2.5}~\cite{codegen2} is a family of auto-regressive language models for program synthesis which are further trained based on CodeGen2~\cite{codegen2} using multilingual programming language data. The variants of this family of models include the standard version {\em multi}, {\em mono} that is further trained based on {\em multi} using additional Python tokens, and {\em instruct} that is further trained based on {\em mono} using instruction data. The size of the parameters is 7B. In our study, we use {\em CodeGen2.5-7B-mono} to run the experiments.
    \item \textbf{StarCoder}~\cite{starcoder} is a code LLM trained from code tokens written in 80+ programming languages and further fine-tuned using Python tokens. The size of parameters is 15.5B.
     \item \textbf{WizardCoder}~\cite{luo2023wizardcoder} is a collection of code LLMs trained by Evol-Instruct. WizardCoder has various released versions based on the model sizes, language specificity, and foundation models. 
     In our work, we use \textit{WizardCoder-Python-7B-V1.0}. 
\end{itemize}

\phead{Evaluation Metrics}. Following previous research on code generation~\cite{codex, codegeex, codegen2, codellama, starcoder, luo2023wizardcoder}, we utilize Pass@k~\cite{codex} to measure the functional correctness of the generated code. 
For each problem, the model generates $n$ code solutions, and $k$ ($k \leq n$) of them are selected for evaluation. 
A problem is considered solved if any of the $k$ samples can pass all tests, and we compute the ratio of solved problems as the final result.
We compute Pass@1 and Pass@10 in the experiments and set $n$ to 15 for efficiency. 
Since we find similar overall trends for Pass@1 and Pass@10, we only report the results of Pass@1 in the paper due to space constraints. 
The results of Pass@10 are archived in our replication package~\cite{replication_package}.

\phead{Implementation Details of Experiments.} 
To maximize the reproducibility, we leverage the popular evaluation framework \textit{Code Generation LM Evaluation Harness}~\cite{starcoder} in our experiments.
We use the perturbed prompts as the input and compare them with the results using original prompts. For the computation of Pass@k, we use the default option which estimates the results following the evaluation of Codex~\cite{codex} and generate 15 samples for each prompt. We set the temperature to 0.2 and the maximum length of generation to 512.

\section{Categories of Perturbations}
\label{sec:categories}

In this section, we discuss the categories of NL perturbations derived based on the process of {\em Stage 1} and {\em Stage 2} in Section~\ref{sec:methodology}. 
We divide the categories into five groups based on the type of operations to apply such perturbations, including {\em Addition}, {\em Deletion}, {\em Editing}, {\em Swap}, and {\em Paraphrasing}.
For each category, we illustrate its description and how we implement the perturbation. 
Table~\ref{tab:categories} provides the original prompt and the perturbed prompt for each category as an example. In the example, ``\_'' refers to a space rather than an underscore.

While illustrating the implementation of perturbations, we refer to the times of perturbations as ``{\sf T}'' (i.e., the perturbation is applied {\sf T} times for each prompt).
 {\sf T} is calculated based on the product of: (1) the number of perturbable elements and (2) a pre-set frequency value (i.e., what percent of the perturbable elements will be perturbed). We set the default frequency value for each category based on our observation on the results and attempt to achieve a balance of naturalness and effectiveness. The detailed frequency of each category can be found in our replication package~\cite{replication_package}.

\begin{table*}
\caption{Categories of Perturbations. 
The IDs of the final categories are marked in \textbf{bold} and the initial categories filtered by the survey are with \st{strikethrough}.}
\vspace{-0.2cm}
\centering
\resizebox{\linewidth}{!} {%
\scalebox{1}{
\begin{tabular}{l|l|l|l}
\hline
\rowcolor[HTML]{EFEFEF}
 \textbf{ID} & \textbf{Name} & \textbf{Example (Original)}  & \textbf{Example (Perturbed)} \\

\hline
\rowcolor[rgb]{0.98,0.98,0.98}
\multicolumn{4}{c}{\textbf{Categories of Addition (A1 - A4)}} \\
\hline
\textbf{A1} & 
Extra Space outside Words &
Write a function to\exampleo{\_}replace ... & 
Write a function to\examplep{\_\_}replace ...
\\
\midrule
\textbf{A2} & 
Extra Space inside Words &
Write a \exampleo{function} to replace ... & 
Write a \examplep{func tion} to replace ...
\\

\midrule
\textbf{A3} & 
Repeated Words &
Write a function \exampleo{to} replace ... & 
Write a function \examplep{to to} replace ...
\\


\midrule
\textbf{A4} & 
Repeated Chars &
Write a \exampleo{function} to replace ... & 
Write a \examplep{funnction} to replace ...
\\

\midrule
\st{A5} & 
Synonym Insertion &
Write a function to \exampleo{find} ... & 
Write a function to \examplep{find discover} ...
\\

\midrule
\st{A6} & 
Attaching URL &
Write a function to replace ... & 
Write a function to replace ... \examplep{https://google.com}
\\

\midrule
\st{A7} & 
Attaching Interrogation Statements &
Write a function to replace ... & 
Write a function to replace ... \examplep{what is what}
\\

\hline
\rowcolor[rgb]{0.98,0.98,0.98}
\multicolumn{4}{c}{\textbf{Categories of Deletion (D1 - D4)}} \\
\hline
\textbf{D1} & 
Char Deletion &
Write a \exampleo{function} to find ... & 
Write a \examplep{funtion} to find ...
\\

\midrule
\textbf{D2} & 
Preposition Deletion &
Write a function \exampleo{to} find ... & 
Write a function\examplep{\_}find ...
\\

\midrule
\textbf{D3} & 
Determiner Deletion &
Write a function to reverse \exampleo{the} string and ... & 
Write a function to reverse\examplep{\_}string and ...
\\

\midrule
\textbf{D4} & 
Space Deletion &
Write \exampleo{a function} to find ... & 
Write \examplep{afunction} to find ...
\\

\hline
\rowcolor[rgb]{0.98,0.98,0.98}
\multicolumn{4}{c}{\textbf{Categories of Editing (E1 - E6)}} \\
\hline
\textbf{E1} & 
Keyboard Typo &
\exampleo{Write} a function to replace ... & 
\examplep{Writr} a function to replace ...
\\

\midrule
\textbf{E2} & 
Extra Capital Letter &
\exampleo{Write} a function to replace ... & 
\examplep{WRite} a function to replace ...
\\

\midrule
\textbf{E3} & 
Grammatical Person Variation &
Write a function that \exampleo{replaces} the ... & 
Write a function that \examplep{replace} the ...
\\

\midrule
\textbf{E4} & 
Active and Passive Voice Variation &
... if numbers \exampleo{are sorted} in ascending order. & 
... if numbers \examplep{sort} in ascending order.
\\

\midrule
\textbf{E5} & 
Word Class Variation &
... check the \exampleo{required} numbers of ..." & 
... check the \examplep{requirement} numbers of ..."
\\

\midrule
\textbf{E6} & 
Synonym Substitution &
Write a function to \exampleo{find} the index of ... & 
Write a function to \examplep{locate} the index of ...
\\

\midrule
\st{E7} & 
Random Char Replacement &
Write a \exampleo{function} to replace ... & 
Write a \examplep{func!ion} to replace ...
\\

\hline
\rowcolor[rgb]{0.98,0.98,0.98}
\multicolumn{4}{c}{\textbf{Categories of Swap (S1 - S2)}} \\
\hline
\textbf{S1} & 
Swap Adjacent Chars &
Write a function to \exampleo{reverse} the string ... & 
Write a function to \examplep{revesre} the string ...
\\

\midrule
\textbf{S2} & 
Swap Adjacent Words &
Write a function \exampleo{to reverse} the string ... & 
Write a function \examplep{reverse to} the string ...
\\

\midrule
\st{S3} & 
Swap Chars Randomly  &
Write a function to \exampleo{reverse} the string ... & 
Write a function to \examplep{rrveese} the string ...
\\

\midrule
\st{S4} & 
Middle Random Swap &
Write a function to \exampleo{reverse} the string ... & 
Write a function to \examplep{rsrveee} the string ...
\\

\midrule
\st{S5} & 
Fully Random Swap &
Write a function to \exampleo{reverse} the string ... & 
Write a function to \examplep{esrveer} the string ...
\\

\hline
\rowcolor[rgb]{0.98,0.98,0.98}
\multicolumn{4}{c}{\textbf{Categories of Paraphrasing (P1 - P2)}} \\
\hline
\textbf{P1} & 
Rephrasing Sentence &
Print even numbers from a list of numbers. & 
Given a list of numbers, print the even numbers.
\\

\midrule
\textbf{P2} & 
Declarative to Interrogative &
Print even numbers from a list of numbers. & 
Can you print even numbers from a list of numbers?
\\


\hline
\hline
\rowcolor[rgb]{0.98,0.98,0.98}
\multicolumn{4}{c}{\textbf{Combinations of Co-occurred Categories (C1 - C3)}} \\
\hline
\textbf{C1} & 
A1 + E1 &
Write a function to \exampleo{check} whether ... & 
\examplep{\_}Write a function to \examplep{chevk} whether ...
\\

\midrule
\textbf{C2} & 
A4 + E1 &
Write a \exampleo{function} to \exampleo{check} whether ... & 
Write a \examplep{functiion} to \examplep{chevk} whether ...
\\
\midrule
\textbf{C3} & 
D1 + E1 &
Write a \exampleo{function} to \exampleo{check} whether ... & 
Write a \examplep{funtion} to \examplep{chevk} whether ...
\\

\bottomrule
\end{tabular}}
}
  \vspace{-0.5cm}
\label{tab:categories}
\end{table*}

\vspace{-0.2cm}
\subsection{Categories of Addition.}
The categories of Addition are related to adding an element (e.g., a character, a word, or a space) to the prompt.

\phead{A1: Extra Space outside Words.} Add an extra space outside the words (i.e., before or after a word).

\noindent{\em \underline{Implementation.}} \tool identifies the locations of word boundaries such as the start and end of the prompt or spaces. \tool then randomly selects {\sf T} locations to add an extra space. We use spaCy~\cite{spacy} to extract English words from the prompt for this and the remaining categories.

\phead{A2: Extra Space inside Words.} Add an extra space inside a word, resulting in the original word being split into two words.

\noindent{\em \underline{Implementation.}}  \tool randomly selects {\sf T} words that are larger than 3 characters and adds a space inside the selected words. \tool only inserts a space between the characters of a word (i.e., does not insert space at the start or end of a word).

\phead{A3: Repeated Words.}  Repeat an existing word in the prompt, especially short words like \textit{``to''}, and \textit{``a''}.

\noindent{\em \underline{Implementation.}}  \tool first identifies short words in the prompt that are not larger than three characters and then randomly selects {\sf T} such words and repeats each selected word once.

\phead{A4: Repeated Chars.}  Repeat a character in the prompt.

\noindent{\em \underline{Implementation.}}  \tool randomly selects {\sf T} words in the prompt and repeats a random character in each selected word.

\vspace{-0.2cm}
\subsection{Categories of Deletion.}
The categories of Deletion are related to the removal of an element in the prompt.

\phead{D1: Char Deletion.} Drop a character from the prompt.

\noindent{\em \underline{Implementation.}}  \tool first identifies the characters that meet the following requirements: (1) reside in a word with at least three characters; (2) not at the start or the end of a word; (3) are lowercase alphabets. \tool then randomly drops {\sf T} such characters from the prompt.

\phead{D2: Preposition Deletion.}  Drop a preposition or a subordinating conjunction (e.g., \textit{``to''}, \textit{``of''}) from the prompt.  

\noindent{\em \underline{Implementation.}}  \tool leverages spaCy~\cite{spacy} to identify the part of speech for each word and then randomly drops {\sf T} prepositions and subordinating conjunctions in the prompt. 

\phead{D3: Determiner Deletion.} Drop a determiner (e.g., \textit{``a''}, \textit{``the''}) from the prompt.

\noindent{\em \underline{Implementation.}} \tool randomly drops {\sf T} prepositions identified by using spaCy~\cite{spacy}.

\phead{D4: Space Deletion.}  Drop a space between the words from the prompt.

\noindent{\em \underline{Implementation.}}  \tool identifies the whitespace characters between the words. Such space can include the space on the keyboard, a newline mark, and a tab mark. \tool then randomly drops {\sf T} spaces from the prompt.

\vspace{-0.2cm}
\subsection{Categories of Editing.}
The categories of Editing are related to editing a character or a word in the prompt.

\phead{E1: Keyboard Typo.} Replace a character in the prompt with its adjacent character on the keyboard.

\noindent{\em \underline{Implementation.}} \tool identifies the lowercase alphabetic characters in the prompt and randomly selects {\sf T} characters. For each selected character, \tool then randomly replaces it with a character that is adjacent on the keyboard. 
The retrieval of adjacent characters is based on our predefined mappings. For example, if the original character is \textit{``r''}, \tool then randomly selects from \textit{``e''}, \textit{``d''}, \textit{``f''}, and \textit{``t''} to replace it. Note that our implementation is based on the standard layout (i.e., QWERTY layout) of keyboard.

\phead{E2: Extra Capital Letter.}  Capitalize a lowercase character .

\noindent{\em \underline{Implementation.}}    \tool randomly selects {\sf T} lowercase alphabetic characters and capitalizes them.

\phead{E3: Grammatical Person Variation.} Convert a verb between its base form and its third person singular.

\noindent{\em \underline{Implementation.}} \tool identifies the verbs in the prompt and randomly selects {\sf T} of them. Then \tool interchanges these verbs between their original and third person singular forms with the help of spaCy~\cite{spacy} and LemmInflect~\cite{LemmInflect}.

\phead{E4: Active and Passive Voice Variation.}  Convert a verb between active voice and passive voice.

\noindent{\em \underline{Implementation.}} 
\tool identifies the verbs in the prompt and randomly selects {\sf T} of them. If the word is in the past participle form, \tool converts it to the active form; and vice versa. 
\tool accomplishes it by spaCy~\cite{spacy} and LemmInflect~\cite{LemmInflect}.

\phead{E5: Word Class Variation.}  Convert a word to its different part of speech.

\noindent{\em \underline{Implementation.}} \tool randomly selects {\sf T} words and then transforms their word classes. \tool converts words randomly among the forms of nouns, verbs, adjectives, and adverbs.

\phead{E6: Synonym Substitution.}  Substitute a word to its synonym.

\noindent{\em \underline{Implementation.}} \tool identifies substantive words (i.e., prepositions, determiners, etc. will be excluded) and randomly selects {\sf T} words. Then \tool replaces these words with their synonyms using WordNet~\cite{fellbaum2010wordnet}, which is widely used by prior studies to retrieve synonyms~\cite{wordnet_in_se,liu2020senet,bhatia2020clustering}.

\vspace{-0.2cm}
\subsection{Categories of Swap.}
The categories of Swap are related to the swap of characters or words in the prompt.

\phead{S1: Swap Adjacent Chars.}  Swap two characters that are adjacent in a word.

\noindent{\em \underline{Implementation.}}  \tool first identifies the characters that meet the following requirements: (1) reside in a word with at least three characters; (2) not at the start or the end of a word; (3) are lowercase alphabetic characters. \tool then randomly selects {\sf T} characters to swap with their next character in the word. While randomly selecting the candidate characters to swap, \tool only selects non-adjacent characters to swap with their next characters to avoid consecutive swaps in the same word.

\phead{S2: Swap Adjacent Words.}  Swap two words that are adjacent in the prompt.

\noindent{\em \underline{Implementation.}}  \tool randomly selects {\sf T} words in the prompt to swap with their next words. Similar to {\sf S1}, \tool only selects non-adjacent words to swap with their next words to avoid consecutive swaps.

\vspace{-0.2cm}
\subsection{Categories of Paraphrasing.}

\phead{P1: Rephrasing Sentence.}  Rephrase the prompt while preserving the original semantic meaning.

\noindent{\em \underline{Implementation.}} \tool splits the prompt into sentences through punctuations and randomly selects {\sf T} sentences. \tool then utilizes paraphraser Parrot~\cite{prithivida2021parrot} to rephrases them.

\phead{P2: Declarative to Interrogative.}  Convert the request in the prompt from declarative to interrogative.

\noindent{\em \underline{Implementation.}} \tool identifies the imperative sentence in the prompt (e.g., ``Write a program...'') and leverages the API of OpenAI LLMs~\cite{openai_api} to convert to an interrogative sentence.

\subsection{Combinations of Co-occurred Categories.}
We discuss the implementation of co-occurred categories summarized from our survey. The details of the survey results can be found in RQ1 of Section~\ref{sec:results}.
The time of perturbations for each member category in the combination is computed independently. It is calculated based on the product of perturbable elements and half of the default frequency value. We refer to the perturbation times of each member category as $T_{1}$ and $T_{2}$.

\phead{C1: Extra Space outside Words \& Keyboard Typo (A1 + E1)}. Add an extra space outside the words and replace a character in the prompt with its adjacent character on the keyboard.

\noindent{\em \underline{Implementation.}} \tool first performs the perturbation of {\em Keyboard Typo} (as discussed in {\sf E1} of this section) $T_{1}$ times. \tool then randomly selects $T_{2}$ locations that are outside of a word to add an extra space.

\phead{C2: Repeated Chars \& Keyboard Typo (A4 + E1)}.  Repeat a character in the prompt and also replace another character with its adjacent character on the keyboard.

\noindent{\em \underline{Implementation.}} \tool randomly selects  $T_{1}$ characters to apply {\em Keyboard Typo} and further randomly selects $T_{2}$ words to repeat a random character in each selected word.

\phead{C3: Char Deletion \& Keyboard Typo (D1 + E1)}.  Drop a character from the prompt and also replace a character with its adjacent character on the keyboard.

\noindent{\em \underline{Implementation.}} After \tool performing {\em Keyboard Typo} $T_{1}$ times, \tool further identifies the characters that are not affected by {\em Keyboard Typo} and randomly drops $T_{2}$ such characters.

\section{Results}
\label{sec:results}
In this section, we discuss the results of each RQ.

\vspace{-0.2cm}
\subsection{RQ1: Practitioners' Perspective on the Categories of Perturbations in NL}

We discuss the results of RQ1 in three aspects: (1) survey participants' perspective for each category; (2) co-occurred categories; and (3) the detailed comments on some categories.

\phead{Survey Participants' Perspective for each Category.} 
Figure~\ref{fig:survey_practice} shows the results of whether the survey participants consider each category of perturbations can occur in real-world scenarios. We find that there are 18 categories of which {\em Yes} is the majority response. The percentage varies from 50.0\% for {\em A3: Repeated Words} and {\em S2: Swap Adjacent Words} to 96.2\% for {\em E6: Synonym Substitution}.

For the remaining 7 categories, the majority response is {\em ``No''}. Among them, {\em E7: Random Char Replacement}, {\em S3: Swap Chars Randomly}, {\em S4: Middle Random Swap}, and {\em S5: Fully Random Swap} have an over 70\% percentage of negative rates. 
These categories are used by prior robustness-related studies~\cite{belinkov2018synthetic,zhuo2023robustness,zhu2023promptbench} for perturbation.

\phead{Co-occurred Categories.} 
Among the 52 valid responses, 45 (86.5\%) participants consider that multiple categories of perturbations can occur in the same prompt, and 7 (13.5\%) participants are not sure of this question.

We received 31 responses regarding the combinations of co-occurred categories. We further filter 2 responses that provide general comments such as {\em ``There are many possible combinations of different perturbations''}. For the remaining 29 responses, the participants provided 144 combinations in total. We then identify the unique combinations and count the frequency. Overall, there are 80 unique combinations provided by the participants. 
We include the three combinations of which frequency is larger than 5 in our study and implement the perturbators correspondingly. 
{\sf C1} - {\sf C3} in Table~\ref{tab:categories} presents these co-occurred categories with examples.

\phead{Comments on Categories.} Besides, we also received comments from the survey participants regarding the details of some categories. For example:

\noindent \textit{``I think repeated words may only happen for short words, like ``to to'' or ``the the'' and repeated char could happen in relatively long words.''}

\noindent \textit{``Char deletion seems not easy to occur in very short words, but can be possible in long words.''}

We take their comments into account while implementing the perturbators. For example, we have leveraged the above two comments in the implementations of {\em Repeated Words (A3)} and {\em Char Deletion (D1)}.

\rqbox{\textbf{Summary:} Among the 25 initial categories of perturbations, there are 18 categories that a majority of the survey participants consider they can occur in real-world scenarios. Our survey participants also provide 80 combinations of co-occurred categories and three of them are mentioned more than five times.}

\begin{figure*}
 \centering
\includegraphics[width=1.00\linewidth]{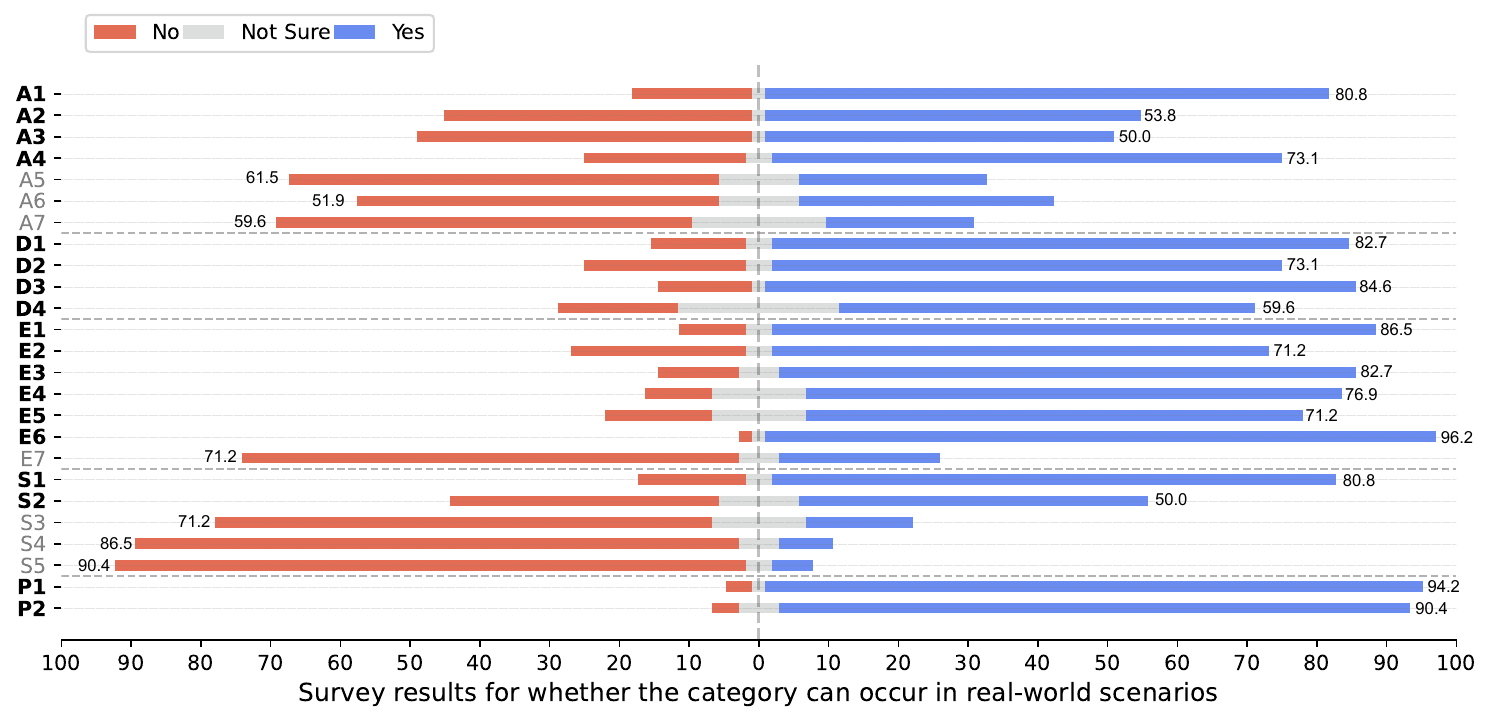}
 \caption{Survey participants' rating for each category (RQ1). Categories in the final list are marked in bold; the majority answer is marked with its number of percentage.}
 \label{fig:survey_practice}
 \vspace{-0.5cm}
 \end{figure*}

\subsection{RQ2: Automatically Composing Perturbations in NL}

\begin{table*}
    \caption{Results of human evaluation on automated perturbations of each category (RQ2). }
    \centering
    
    \tabcolsep=3pt
\resizebox{\linewidth}{!} {
    \begin{tabular}{l|llll|llll|llllll|ll|ll|lll|l}
        \hline
        \rowcolor[HTML]{EFEFEF}
         Dataset & \textbf{A1} &  \textbf{A2}  &\textbf{A3}  &\textbf{A4}      &\textbf{D1} &\textbf{D2} &\textbf{D3} &\textbf{D4}    &\textbf{E1} &\textbf{E2} &\textbf{E3} &\textbf{E4} &\textbf{E5} &\textbf{E6}   &\textbf{S1} &\textbf{S2}   &\textbf{P1} &\textbf{P2}        &\textbf{C1} &\textbf{C2} &\textbf{C3}     &\textit{Avg.} \\
        \hline
        \rowcolor[rgb]{0.98,0.98,0.98}
        \multicolumn{23}{c}{\textbf{Effectiveness (\%)}} \\
        \hline
        HumanEval & 100.0   & 98.2 & 99.4  & 100.0          & 99.4 & 99.4 & 97.6 & 98.8        & 97.0   & 97.6 & 89.0  & 82.3 & 93.9 & 87.8         & 97.6 & 97.6        & 51.2   & 97.6        & 97.0   & 98.8   & 95.7     & 94.0 \\ 
        MBPP & 100.0   & 99.8 & 99.3  & 98.8          & 100.0 & 93.4 & 100.0 & 100.0        & 99.8   & 99.8 & 99.5  & 99.5 & 99.1 & 73.5         & 98.6 & 98.1        & 80.8   & 96.7        & 99.8   & 99.5   & 100.0       & 96.7\\

        \hline
        \rowcolor[rgb]{0.98,0.98,0.98}
        \multicolumn{23}{c}{\textbf{Naturalness (\%)}} \\
        \hline
        HumanEval  & 100.0   & 97.6 & 92.1  & 92.1         & 98.8 & 98.2 & 98.8 & 97.6        & 90.9   & 98.2 & 98.8  & 67.1 & 79.3 & 21.3         & 97.0 & 68.3        & 85.4   & 60.4        & 91.5   & 92.7   & 94.5      & 86.6 \\ 
        MBPP   & 100.0   & 98.1 & 99.5  & 99.1          & 99.5 & 98.8 & 100.0 & 100.0        & 94.6   & 98.6 & 99.5  & 67.9 & 90.6 & 41.2         & 99.8 & 76.4       & 81.0  & 94.2        & 96.5   & 96.3   & 95.8      & 91.8 \\ 
        \hline
    \end{tabular}
    }

    \vspace{-0.3cm}

    \label{tab:rq2}

\end{table*}

In this RQ, we present the results of our human evaluation on the perturbed prompts. As discussed in Section~\ref{sec:methodology}, we examine the effectiveness (i.e., if the perturbation belongs to the desired category and is applied to the NL part of the prompt) and naturalness (i.e., if the resulted prompt looks natural) of each perturbed prompt. 
We study the 18 final categories and 3 combinations of co-occurred categories. 
For the result that is either not effective or not natural, we manually modify it to be effective and natural.
We then have two datasets with manually verified perturbations of each category based on HumanEval and MBPP that can be used to examine the robustness of LLMs on code generation, namely {\sf HumanEval-R} and {\sf MBPP-R}.

\phead{Results.}
Table~\ref{tab:rq2} presents the results of this RQ.
Overall, \tool has an effectiveness and naturalness of over 90\% for 14 out of the 21 categories for both of the two datasets. For HumanEval, the effectiveness ranges from 51.2\% of {\em Rephrasing Sentence (P1)} to 100.0\% of {\em Repeated Chars (A4)}, the naturalness ranges from 21.3\% of {\em Synonym Substitution (E6)} to 100\% of {\em Extra Space outside Words (A1)}. For MBPP, the effectiveness ranges from 73.5\% of {\em Synonym Substitution (E6)} to 100\% of {\em Char Deletion (D1)}, the naturalness ranges from 41.2\% of {\em Synonym Substitution (E6)} to 100\% of {\em Determiner Deletion (D3)}.  \tool has an average effectiveness and naturalness of 94.0\% \& 86.6\% for HumanEval, and 96.7\% \& 91.8\% for MBPP, respectively.

\phead{Case Study.}
We discuss the cases of perturbations in four aspects: \ding{182} both effective and natural; \ding{183} effective but not natural; \ding{184} not effective but natural; \ding{185} not effective and not natural. 
As showns in Figure~\ref{fig:rq2_case_study}, we provide two perturbed examples for \ding{182} and one pair of examples each for \ding{183}\ding{184}\ding{185}, respectively. Each pair of the examples for \ding{183}\ding{184}\ding{185} include an automatically perturbed prompt on the left and a manually modified prompt on the right. We only present the NL part in the prompt with differences for brevity.

\noindent{\ding{182} \em \underline{Effective \& Natural.}} 
As shown in example 1 of Figure~\ref{fig:rq2_case_study}, the first sentence in the prompt is transformed from a declarative sentence (i.e., \textit{``For a given list...''}) to an interrogative sentence (i.e., \textit{``Can you calculate Mean...''}) by \tool. The change is consistent with the requirements of the perturbation category \textit{P2: Declarative to Interrogative} and the resulting prompt looks natural.
For example 2, the automated perturbator of \textit{E1: Keyboard Typo} changes the letter \textit{``r''} in the word \textit{``cha\textbf{r}acters''} to the adjacent character \textit{``t''}, which matches the definition of its category. 
Besides, it seems natural to the scenario of inputting adjacent characters by accident when using the keyboard.

\noindent{\ding{183} \em \underline{Effective \& Not Natural.}} 
On the left side of this case, the word \textit{``odd''} is automatically perturbated to \textit{``queer''} for the category \textit{E6: Synonym Substitution}. 
Although the perturbator effectively replaces the word with one of the synonyms of \textit{``odd''}, it does not understand the actual meaning of \textit{odd} in the context (i.e., denoting the opposite of even numbers). Therefore, the resulting synonym is unnatural given the context of this prompt.
The process of manual verification replaces the word \textit{``product''} with \textit{``multiplication''} which remains the same meaning.

\noindent{\ding{184} \em \underline{Not Effective \& Natural.}} 
In this example, the perturbator is for the category of \textit{P1: Rephrasing Sentence}. However, the perturbator simply removes the word \textit{``the''} and the period at the end instead of actually paraphrasing the sentence. Therefore, although the resulting prompt looks natural, it does not effectively follow the definition of its category.
We manually change the verb (i.e., \textit{``Write''} → \textit{``Develop''}) and structure (i.e., \textit{``to find''} → \textit{``that calculates''}) to rephrase the prompt and preserve the same semantic meaning.

\noindent{\ding{185} \em \underline{Not Effective \& Not Natural.}} 
For \textit{S2: Swap Adjacent Words}, the perturbator is supposed to swap two adjacent words in the sentence. However, in this example, the word connected by the hyphen \textit{``non-prime''} is separated, and the prefix \textit{``non''} is exchanged with the adjacent word \textit{``identify''} due to the tokenizer's identification of different words. Therefore, the perturbed prompt is neither effective nor natural.
To have an effective and natural perturbed prompt, we manually swap word \textit{``identity''} with \textit{``to''} as shown on the right of this example.

\rqbox{\textbf{Summary:} \tool has an effectiveness \& naturalness of over 90\% for 14 out of the 21 categories for both datasets, with an average effectiveness \& naturalness of 94.0\% \& 86.6\% for HumanEval, and 96.7\% \& 91.8\% for MBPP, respectively.}

\begin{figure}
 \vspace{-0.2cm}
 \centering
\includegraphics[width=1.00\linewidth]{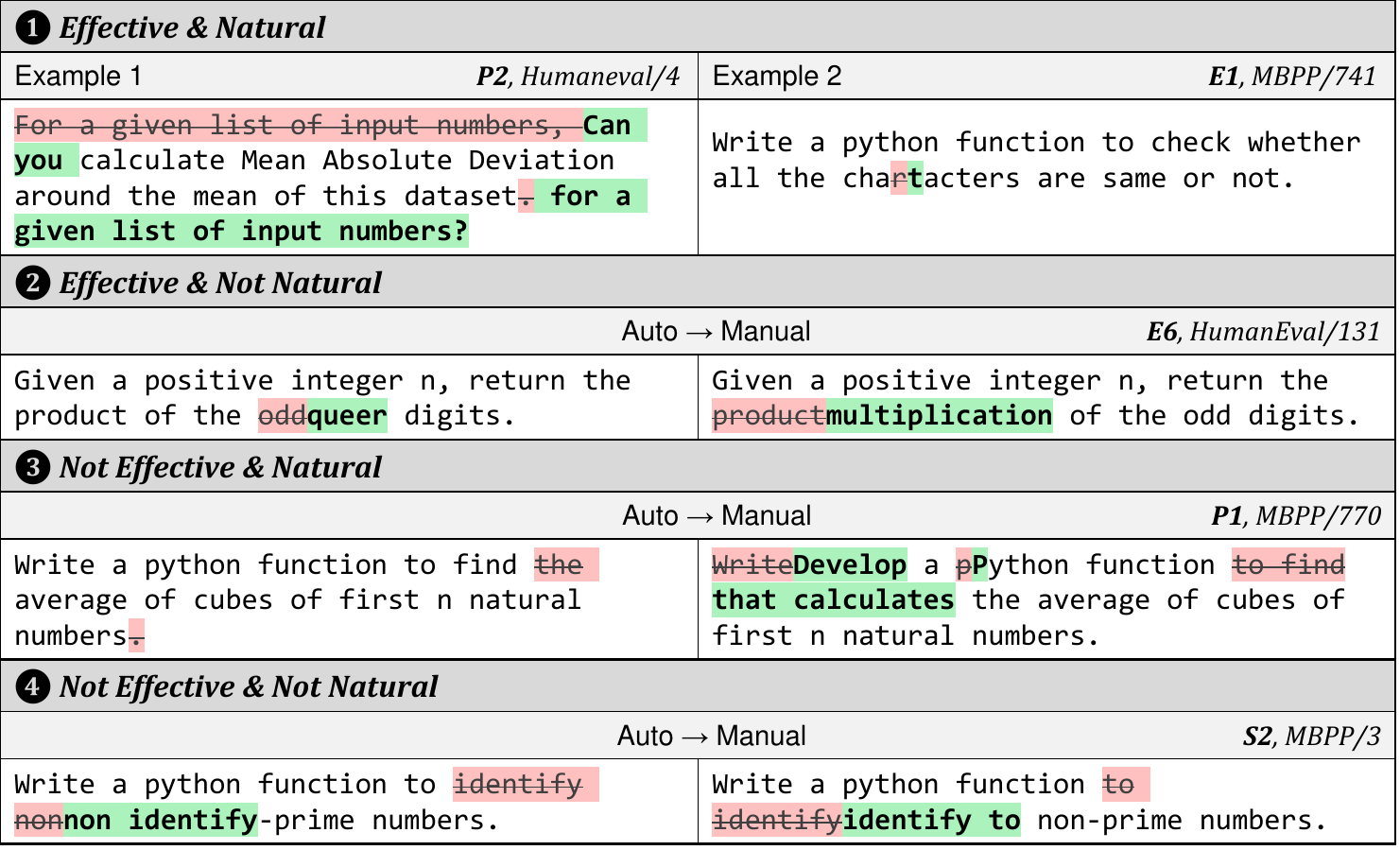}
 \caption{Examples of effectiveness and naturalness in the perturbation (RQ2). }
 \label{fig:rq2_case_study}
 \vspace{-0.2cm}
 \end{figure}

\subsection{RQ3: Robustness of Code LLMs to NL Perturbations on Code Generation}

\begin{table*}
    \caption{Results of Code Generation using the Original Prompt (Ori.) and each Category of the Perturbed Prompts (RQ3). The relative decrease over 3\% compared to Ori. is marked in \down{italic and blue}, and over 5\% is marked in \sdown{italic, bold, and blue}. Avg.: average number of perturbed results.}
     \vspace{-0.2cm}
    \centering
    \tabcolsep=3pt
\resizebox{\linewidth}{!} {
\begin{tabular}{l|c|cccc|cccc|cccccc|cc|cc|ccc|c} 
\hline
{\cellcolor[rgb]{0.937,0.937,0.937}}Model & {\cellcolor[rgb]{0.937,0.937,0.937}}Ori. & {\cellcolor[rgb]{0.937,0.937,0.937}}\textbf{A1} & {\cellcolor[rgb]{0.937,0.937,0.937}}\textbf{A2} & {\cellcolor[rgb]{0.937,0.937,0.937}}\textbf{A3} & {\cellcolor[rgb]{0.937,0.937,0.937}}\textbf{A4} & {\cellcolor[rgb]{0.937,0.937,0.937}}\textbf{D1} & {\cellcolor[rgb]{0.937,0.937,0.937}}\textbf{D2} & {\cellcolor[rgb]{0.937,0.937,0.937}}\textbf{D3} & {\cellcolor[rgb]{0.937,0.937,0.937}}\textbf{D4} & {\cellcolor[rgb]{0.937,0.937,0.937}}\textbf{E1} & {\cellcolor[rgb]{0.937,0.937,0.937}}\textbf{E2} & {\cellcolor[rgb]{0.937,0.937,0.937}}\textbf{E3} & {\cellcolor[rgb]{0.937,0.937,0.937}}\textbf{E4} & {\cellcolor[rgb]{0.937,0.937,0.937}}\textbf{E5} & {\cellcolor[rgb]{0.937,0.937,0.937}}\textbf{E6} & {\cellcolor[rgb]{0.937,0.937,0.937}}\textbf{S1} & {\cellcolor[rgb]{0.937,0.937,0.937}}\textbf{S2} & {\cellcolor[rgb]{0.937,0.937,0.937}}\textbf{P1} & {\cellcolor[rgb]{0.937,0.937,0.937}}\textbf{P2} & {\cellcolor[rgb]{0.937,0.937,0.937}}\textbf{C1} & {\cellcolor[rgb]{0.937,0.937,0.937}}\textbf{C2} & {\cellcolor[rgb]{0.937,0.937,0.937}}\textbf{C3} & {\cellcolor[rgb]{0.937,0.937,0.937}}\textit{\textbf{Avg.}}  \\ 
\hline
\multicolumn{24}{c}{{\cellcolor[rgb]{0.98,0.98,0.98}}\cellcolor[rgb]{0.98,0.98,0.98}\textbf{HumanEval (Pass@1)}} \\ 
\hline
InCoder                                   & 16.1                                     & 15.9                                            & \sdown{13.4}                   & 16.0                                            & \sdown{15.2}                   & \down{15.7}                    & 16.3                                            & 16.0                                            & \sdown{14.6}                   & \sdown{13.6}                   & 16.2                                            & \down{15.7}                    & \down{15.4}                    & \down{15.7}                    & \down{15.6}                    & \sdown{15.1}                   & \down{15.5}                    & \sdown{13.9}                   & \sdown{15.3}                   & \sdown{14.9}                   & \sdown{15.1}                   & \sdown{14.6}                                       & \sdown{15.2}  \\

Code Llama                                & 30.5                                     & 30.0                                            & \sdown{28.5}                   & \down{29.1}                    & \down{29.4}                    & 30.3                                            & \sdown{28.7}                   & \sdown{28.8}                   & \down{29.3}                    & 29.6                                            & \down{29.2}                    & 29.7                                            & 30.9                                            & 29.6                                            & \down{29.2}                    & 29.7                                            & \down{29.3}                    & \sdown{28.6}                   & 29.8                                            & 30.0                                            & \down{29.4}                    & \sdown{28.8}                                       & \down{29.4}   \\

CodeGeeX2                                 & 33.2                                     & \down{31.8}                    & \sdown{30.6}                   & \down{31.6}                    & \down{32.1}                    & \sdown{31.4}                   & 32.6                                            & \down{32.2}                    & \sdown{31.0}                   & \sdown{31.4}                   & 32.3                                            & 32.8                                            & 32.5                                            & \down{31.9}                    & \down{32.2}                    & \down{31.9}                    & 32.3                                            & \sdown{30.9}                   & \sdown{31.4}                   & 32.8                                            & 32.6                                            & \down{31.7}                                        & \down{31.9}   \\

CodeGen2.5                                & 33.4                                     & 33.4                                            & 32.5                                            & \down{32.3}                    & \sdown{31.0}                   & \sdown{31.3}                   & 32.5                                            & 32.9                                            & 32.6                                            & \down{32.2}                    & \down{32.3}                    & 33.2                                            & 33.1                                            & 32.8                                            & 32.9                                            & \sdown{31.5}                   & 33.4                                            & \down{32.1}                    & \sdown{30.2}                   & \down{32.4}                    & \sdown{31.7}                   & \sdown{31.1}                                       & \down{32.2}   \\

StarCoder                                 & 33.9                                     & 33.0                                            & 33.2                                            & 33.9                                            & 33.5                                            & \down{32.9}                    & \down{32.9}                    & 33.4                                            & 33.7                                            & \sdown{32.0}                   & 34.4                                            & 33.1                                            & 34.2                                            & 33.4                                            & 33.9                                            & 34.0                                            & 33.2                                            & 33.2                                            & \sdown{31.3}                   & 33.7                                            & 33.6                                            & 33.2                                                                & 33.3                           \\

WizardCoder                               & 43.9                                     & 43.5                                            & \down{42.1}                    & 42.8                                            & 43.8                                            & 42.9                                            & 43.2                                            & 43.9                                            & \down{42.0}                    & \down{41.7}                    & 44.4                                            & 43.5                                            & 43.8                                            & 44.3                                            & 43.3                                            & 42.8                                            & 43.3                                            & \sdown{41.1}                   & 44.0                                            & 44.7                                            & 43.1                                            & \sdown{41.5}                                       & 43.1                           \\

\hline
\textit{Avg.}                             & 31.8                                     & 31.3                                            & \sdown{30.0}                   & 31.0                                            & \down{30.8}                    & \down{30.7}                    & \down{31.0}                    & 31.2                                            & \down{30.5}                    & \sdown{30.1}                   & 31.5                                            & 31.3                                            & 31.7                                            & 31.3                                            & 18.3                                            & \down{30.8}                    & 31.2                                            & \sdown{30.0}                   & \down{30.3}                    & 31.4                                            & 30.9                                            & \sdown{30.2}                                       & \down{30.9}   \\

\hline
\multicolumn{24}{c}{{\cellcolor[rgb]{0.98,0.98,0.98}}\cellcolor[rgb]{0.98,0.98,0.98}\textbf{MBPP (Pass@1)}} \\ 
\hline
InCoder                                   & 19.7                                     & \sdown{17.5}                   & \down{19.0}                    & 20.3                                            & 19.7                                            & \sdown{18.3}                   & 19.2                                            & 20.0                                            & \sdown{19.0}& \sdown{18.4}                   & 19.6                                            & 20.7                                            & 19.7                                            & 19.4                                            & 19.9                                            & 19.1                                            & 19.9                                            & 19.8                                            & \sdown{15.6}                   & 19.3                                            & 19.5                                            & 19.1                                                                & 19.2\\

Code Llama                                & 49.6                                     & 50.2                                            & 48.9                                            & 49.0                                            & \down{48.0}                    & 48.4                                            & 48.2                                            & 49.9                                            & 48.8& \down{47.8}                    & 50.7                                            & \down{48.1}                    & \down{48.0}                    & \down{47.3}                    & 49.0                                            & 48.4                                            & 49.4                                            & 48.5                                            & \sdown{46.0}                   & 48.5                                            & \sdown{46.7}                   & \down{47.9}                                        & 48.5\\

CodeGeeX2                                 & 45.8                                     & 45.9                                            & 45.3                                            & 45.9                                            & 45.5                                            & 45.6                                            & 44.8                                            & 45.6                                            & 46.1& \down{43.7}                    & 47.0                                            & \down{43.6}                    & \sdown{41.8}                   & 44.5                                            & \down{44.2}                    & 45.7                                            & \sdown{43.1}                   & 45.2                                            & \sdown{41.7}                   & 44.5                                            & 44.8                                            & \down{43.6}                                        & 44.7\\

CodeGen2.5                                & 54.3                                     & 54.5                                            & 53.9                                            & 54.4                                            & 53.8                                            & 54.2                                            & \down{52.6}                    & 54.1                                            & 54.0                                            & 53.8                                            & 53.1                                            & \down{52.6}                    & \down{51.7}                    & 53.1                                            & 53.0                                            & 54.2                                            & 54.5                                            & 53.9                                            & \down{51.6}                    & 52.9                                            & 52.9                                            & \down{52.3}                                        & 53.4                           \\

StarCoder                                 & 56.0                                     & 55.5                                            & 55.5                                            & 55.7                                            & 55.6                                            & 54.6                                            & 54.9                                            & 56.0                                            & 56.0& \down{54.3}                    & 55.5                                            & \down{53.9}                    & \down{53.9}                    & 55.8                                            & 54.8                                            & 55.6                                            & 56.5                                            & 55.8                                            & 54.5                                            & 55.6                                            & \down{54.2}                    & 55.4                                                                & 55.2                           \\

WizardCoder                               & 57.1                                     & 56.2                                            & 56.9                                            & 56.6                                            & 56.1                                            & 56.4                                            & 55.8                                            & 56.5                                            & 57.0& 56.3                                            & 56.8                                            & 56.9                                            & 56.6                                            & 57.2                                            & 56.6                                            & 57.4                                            & 55.5                                            & \down{55.0}                    & 56.1                                            & 56.2                                            & 56.1                                            & \down{54.6}                                        & 56.3                           \\ 
\hline
\textit{Avg.}                             & 47.1                                     & 46.7                                            & 46.6                                            & 47.0                                            & 46.4                                            & 46.3                                            & 45.9                                            & 47.0                                            & 46.8& \down{45.7}                    & 47.1                                            & 46.0                                            & \down{45.3}                    & 46.2                                            & 27.7                                            & 46.7                                            & 46.5                                            & 46.4                                            & \sdown{44.2}                   & 46.2                                            & \down{45.7}                    & \down{45.5}                                        & 46.2                           \\
\hline
\end{tabular}
}
\label{tab:rq3-1}
 \vspace{-0.3cm}
\end{table*}

In this research question, we discuss the results of code generation using original prompts and the perturbed results. We use the six LLMs discussed in Section~\ref{sec:methodology} to run the experiments of code generation.  We run the experiments of code generation on our manually verified dataset with default frequency (i.e., {\sf HumanEval-R} and {\sf MBPP-R}) and compare them with the results of using original prompts.

\phead{Results.}
Table~\ref{tab:rq3-1} presents the results of this sub-RQ. The numbers of relative decrease over 3\% and 5\% are particularly marked. Overall, we find that all the categories of perturbations lead to decreases in the results for the six code LLMs in general. The extent of decrease varies for different models and different categories.

\noindent{\em \underline{Results for Different Categories.}} 
We find that the perturbations of some categories have more considerable decreases in the performance of code generation. For example, {\em Declarative to Interrogative (P2)} decreases the Pass@1 of HumanEval and MBPP datasets to 30.3\% (relatively decrease by 4.8\%) and 44.2\% (relatively decrease by 6.1\%) on average, respectively. Particularly, the perturbation of {\em P2} decreases the Pass@1 of Incoder in MBPP by 21.2\% relatively, from 19.7\% to 15.6\%, which is the largest decrease among all the results. In contrast, categories such as {\em Extra Space outside Words (A1)} decrease the results by a relatively small margin (i.e., relative decreases of 1.7\% and 0.9\% on average for HumanEval and MBPP, respectively).

\noindent{\em \underline{Results for Different Models.}} 
The last column of Table~\ref{tab:rq3-1} shows the average results of each model, excluding the results of using original prompts. We find that there is a decrease of Pass@1 for all the six code LLMs in general. Among the models, Incoder has a more noticeable decrease (e.g., a relative decrease of 5.6\% for HumanEval, 15.2\% {\em vs.} 16.1\%) while StarCoder and WizardCoder have a relatively mild decrease. 
The potential reason might be that StarCoder has a larger size of parameters than other models (e.g., 15.5B {\em vs.} 6-7B), and WizardCoder is further fine-tuned using Python code, which may enhance the overall capability of LLMs in comprehending various scenarios and connecting NL with python code.

\begin{figure}
 \centering
\includegraphics[width=0.97\linewidth]{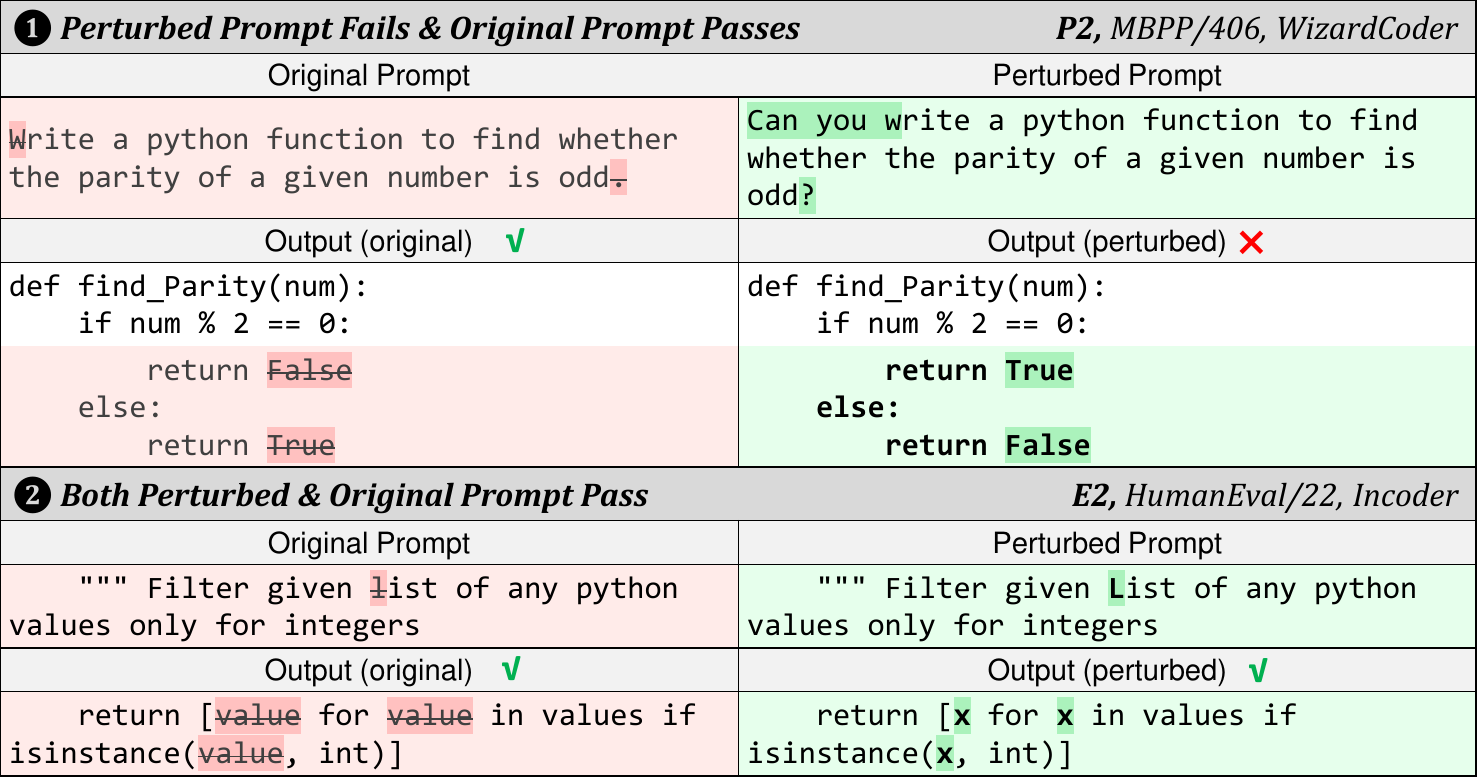}
 \caption{Examples of code generation using our manually-verified perturbation datasets (RQ3).  We only present the part with differences inside each pair of prompts and generated code for brevity.}
 \label{fig:rq3-1_case_study}
 \vspace{-0.2cm}
 \end{figure}

\phead{Case Study.} We discuss the cases in two aspects: the generated code using \ding{182} the perturbed prompt fails the tests but the original prompt passes the tests, and the generated code using \ding{183} perturbed and original prompts both pass the tests. Figure~\ref{fig:rq3-1_case_study} shows the examples of generated code with and without perturbations.

\noindent{\ding{182} \em \underline{Perturbed Prompt Fails \& Original Prompt Passes.}}
As shown in Figure~\ref{fig:rq3-1_case_study}, the natural language part of the prompt is converted from a declarative sentence \textit{``Write ...is odd.''} to an interrogative sentence \textit{``Can you ...?''} following the category of {\em Declarative to Interrogative (P2)}. 
However, with such limited modifications, the logic of the code generated by the model is completely opposite (i.e., an interchange between return values True and False), consequently leading to the failure in this code generation request (i.e., judging parity). 
Despite that \textit{WizardCoder} is relatively less affected by other perturbations (i.e., as shown in the results of Table~\ref{tab:rq3-1}), the changes of sentence structure (although the same semantic meaning is preserved) may still impact LLMs for their understanding of the prompt.

\noindent{\ding{183} \em \underline{Both Perturbed \& Original Prompt Pass.}} 
In this case, the first letter \textit{``l''} of word \textit{``list''} is capitalized (i.e., ``\textit{\textbf{L}ist}'') in the prompt following the category of {\em Extra Capital Letter (E2)}. 
By comparing the generated code before and after perturbations, although there are differences in the name of identifiers within the list (i.e., \textit{``value''} → \textit{``x''}), the semantics remain the same. 
The results in Table~\ref{tab:rq3-1} and this example suggest that these LLMs might be relatively insensitive to the case of natural language in the prompt.

\rqbox{\textbf{Summary:} The performance of code generation is decreased by the perturbed prompts, while the extent varies for different categories and code LLMs. }

\section{Discussion}
\label{sec:discussion}

\vspace{-0.1cm}
\subsection{Implications}

\vspace{-0.1cm}
\phead{Implications for Prompt Engineering.}
(1) As shown in the results of RQ3, the perturbations of NL description in the prompts can decrease the performance of code generation using LLMs. A variation of one word in the prompts may result in the returned code solution from correct to incorrect. Our survey participants consider that such perturbations may often happen in real-world scenarios. Therefore, \textbf{practitioners should be attentive while composing the prompts}. For example, carefully verifying the syntax of the prompts to avoid typos and grammar issues by integrating a syntax detector or applying specific pre-processing strategies. (2) We also find that transforming the prompts from declarative (e.g., \textit{``Write a Python function that ...''}) to interrogative (e.g., \textit{``Can you write a python function that ...''}) can noticeably decrease the performance of code generation. \textbf{Declarative statements might be more suggested to use for the description of code generation by LLMs}. 
(3) We also find that there is a more considerable overall decrease in HumanEval than MBPP. The potential reason might be that the average number of words in NL descriptions of Humaneval is much larger than MBPP (i.e., 47.0 vs. 17.9, as shown in Table~\ref{tab:datasets}), and the perturbations may have a more obvious impact on datasets with more complicated descriptions. Therefore, \textbf{for more sophisticated requirements of code generation, a more concrete and precise NL may be needed}.
Overall, our findings highlight the importance of constructing proper prompts to better leverage LLMs in SE tasks such as code generation.

\phead{Implications for the Robustness of LLMs.}
In our survey, practitioners consider that the perturbations of NL may occur in real-world scenarios when using LLMs to perform code generation. For example, a participant commented that \textit{``People may have bad command of English, so some grammatical variations may be acceptable''}. The participant mentioned that users are not always writing perfect English, LLMs are preferred to also provide coherent results given prompts with grammatical variations. Therefore, \textbf{the robustness to real-world variations in NL is an important aspect of the practicability and user experience of LLMs in code generation}.
In RQ3, we find that the \textbf{current code LLMs are not sufficiently robust to the prompts with these perturbations}. 
Future studies may consider investigating how to improve such robustness and build more practically useful LLMs in the real world.

\vspace{-0.2cm}
\subsection{Threats to Validity}
\vspace{-0.1cm}

\phead{Internal Validity.}
We manually evaluate the perturbed prompts for their effectiveness and naturalness. To mitigate the subjectivity in this process, the two authors of this paper follow the process of prior studies~\cite{huo2023autolog,dlfinder} and independently label the results and discuss each disagreement until a consensus is reached. We have a substantial agreement during this process, with a Cohen's Kappa value of 0.74.
We summarize 25 initial perturbation categories based on literature review and practical experiences. The summarized categories may not be complete and reflect all the possible scenarios. However, the implementation of each perturbator in \tool is independent, which makes \tool easy to extend and include additional perturbators.

\phead{External Validity.}
We only include open-source code LLMs in our experiments and we do not study commercial LLMs (e.g., ChatGPT~\cite{chatgpt}). 
Our concern is that commercial LLMs are continuously updating and evolving, the studied version may be outdated and permanently not available anymore. In contrast, open-source LLMs can be stably accessed over time once they are released and the historical versions can be retrospected.
We choose code LLMs to conduct the study as they can directly perform code generation on-the-fly, without the need of fine-tuning the models using additional code data. Future research may study the robustness in real-world scenarios on other language models.
We conduct the study on the task of code generation only. However, our perturbation technique on NL is general. Future studies may consider verifying the findings of our study on other software engineering tasks.

\section{Conclusion and Future Work}
\label{sec:conclusion}

In this paper, we study and explore the robustness of code LLMs to NL perturbations in real-world scenarios. 
First, we derive 18 perturbation categories that may happen in the real world through a literature review and a comprehensive survey with practitioners. 
Based on these categories, we then perform automated perturbation and manual validation on two widely used code generation datasets. 
Finally, we evaluate the robustness of six open-source code LLMs on the manually verified datasets.
Our study provides practitioners and researchers with a practical perspective on the robustness of code LLMs and sheds light on future directions for prompt engineering and comprehensively improving the capability of LLMs.

\balance
\bibliographystyle{IEEEtran}
\bibliography{paper}

\begin{thebibliography}{10}
\providecommand{\url}[1]{#1}
\csname url@samestyle\endcsname
\providecommand{\newblock}{\relax}
\providecommand{\bibinfo}[2]{#2}
\providecommand{\BIBentrySTDinterwordspacing}{\spaceskip=0pt\relax}
\providecommand{\BIBentryALTinterwordstretchfactor}{4}
\providecommand{\BIBentryALTinterwordspacing}{\spaceskip=\fontdimen2\font plus
\BIBentryALTinterwordstretchfactor\fontdimen3\font minus \fontdimen4\font\relax}
\providecommand{\BIBforeignlanguage}[2]{{%
\expandafter\ifx\csname l@#1\endcsname\relax
\typeout{** WARNING: IEEEtran.bst: No hyphenation pattern has been}%
\typeout{** loaded for the language `#1'. Using the pattern for}%
\typeout{** the default language instead.}%
\else
\language=\csname l@#1\endcsname
\fi
#2}}
\providecommand{\BIBdecl}{\relax}
\BIBdecl

\bibitem{liu2023fill}
Z.~Liu, C.~Chen, J.~Wang, X.~Che, Y.~Huang, J.~Hu, and Q.~Wang, ``Fill in the blank: Context-aware automated text input generation for mobile gui testing,'' in \emph{2023 IEEE/ACM 45th International Conference on Software Engineering (ICSE)}.\hskip 1em plus 0.5em minus 0.4em\relax IEEE, 2023, pp. 1355--1367.

\bibitem{feng2023prompting}
S.~Feng and C.~Chen, ``Prompting is all your need: Automated android bug replay with large language models,'' in \emph{IEEE/ACM 46th International Conference on Software Engineering (ICSE)}.\hskip 1em plus 0.5em minus 0.4em\relax IEEE, 2024.

\bibitem{geng2023large}
M.~Geng, S.~Wang, D.~Dong, H.~Wang, G.~Li, Z.~Jin, X.~Mao, and X.~Liao, ``Large language models are few-shot summarizers: Multi-intent comment generation via in-context learning,'' in \emph{IEEE/ACM 46th International Conference on Software Engineering (ICSE)}.\hskip 1em plus 0.5em minus 0.4em\relax IEEE, 2024.

\bibitem{codex}
M.~Chen, J.~Tworek, H.~Jun, Q.~Yuan, H.~P. de~Oliveira~Pinto, J.~Kaplan, H.~Edwards, Y.~Burda, N.~Joseph, G.~Brockman, A.~Ray, R.~Puri, G.~Krueger, M.~Petrov, H.~Khlaaf, G.~Sastry, P.~Mishkin, B.~Chan, S.~Gray, N.~Ryder, M.~Pavlov, A.~Power, L.~Kaiser, M.~Bavarian, C.~Winter, P.~Tillet, F.~P. Such, D.~Cummings, M.~Plappert, F.~Chantzis, E.~Barnes, A.~Herbert-Voss, W.~H. Guss, A.~Nichol, A.~Paino, N.~Tezak, J.~Tang, I.~Babuschkin, S.~Balaji, S.~Jain, W.~Saunders, C.~Hesse, A.~N. Carr, J.~Leike, J.~Achiam, V.~Misra, E.~Morikawa, A.~Radford, M.~Knight, M.~Brundage, M.~Murati, K.~Mayer, P.~Welinder, B.~McGrew, D.~Amodei, S.~McCandlish, I.~Sutskever, and W.~Zaremba, ``Evaluating large language models trained on code,'' \emph{arXiv preprint arXiv:2107.03374}, 2021.

\bibitem{starcoder}
R.~Li, L.~B. Allal, Y.~Zi, N.~Muennighoff, D.~Kocetkov, C.~Mou, M.~Marone, C.~Akiki, J.~Li, J.~Chim \emph{et~al.}, ``Starcoder: may the source be with you!'' \emph{arXiv preprint arXiv:2305.06161}, 2023.

\bibitem{llama2}
H.~Touvron, L.~Martin, K.~Stone, P.~Albert, A.~Almahairi, Y.~Babaei, N.~Bashlykov, S.~Batra, P.~Bhargava, S.~Bhosale \emph{et~al.}, ``Llama 2: Open foundation and fine-tuned chat models,'' \emph{arXiv preprint arXiv:2307.09288}, 2023.

\bibitem{lin2024llm}
F.~Lin, D.~J. Kim \emph{et~al.}, ``When llm-based code generation meets the software development process,'' \emph{arXiv preprint arXiv:2403.15852}, 2024.

\bibitem{icse24_code_suggestion_rag}
J.~Chen, X.~Hu, Z.~Li, C.~Gao, X.~Xia, and D.~Lo, ``Code search is all you need? improving code suggestions with code search,'' in \emph{Proceedings of the 46th {IEEE/ACM} International Conference on Software Engineering, {ICSE} 2024, Lisbon, Portugal, April 14-20, 2024}, 2024, pp. 73:1--73:13.

\bibitem{shen2023chatgpt}
X.~Shen, Z.~Chen, M.~Backes, and Y.~Zhang, ``In chatgpt we trust? measuring and characterizing the reliability of chatgpt,'' \emph{arXiv preprint arXiv:2304.08979}, 2023.

\bibitem{wang2023robustness}
J.~Wang, H.~Xixu, W.~Hou, H.~Chen, R.~Zheng, Y.~Wang, L.~Yang, W.~Ye, H.~Huang, X.~Geng \emph{et~al.}, ``On the robustness of chatgpt: An adversarial and out-of-distribution perspective,'' in \emph{ICLR 2023 Workshop on Trustworthy and Reliable Large-Scale Machine Learning Models}, 2023.

\bibitem{wang2022adversarial}
B.~Wang, C.~Xu, S.~Wang, Z.~Gan, Y.~Cheng, J.~Gao, A.~H. Awadallah, and B.~Li, ``Adversarial glue: A multi-task benchmark for robustness evaluation of language models,'' 2022.

\bibitem{zhu2023promptbench}
K.~Zhu, J.~Wang, J.~Zhou, Z.~Wang, H.~Chen, Y.~Wang, L.~Yang, W.~Ye, N.~Z. Gong, Y.~Zhang, and X.~Xie, ``Promptbench: Towards evaluating the robustness of large language models on adversarial prompts,'' 2023.

\bibitem{chatgpt}
``{ChatGPT}, {OpenAI},'' \url{https://chat.openai.com}, 2023, last accessed August 2023.

\bibitem{ye2023assessing}
W.~Ye, M.~Ou, T.~Li, X.~Ma, Y.~Yanggong, S.~Wu, J.~Fu, G.~Chen, J.~Zhao \emph{et~al.}, ``Assessing hidden risks of llms: An empirical study on robustness, consistency, and credibility,'' \emph{arXiv preprint arXiv:2305.10235}, 2023.

\bibitem{wang2023recode}
\BIBentryALTinterwordspacing
S.~Wang, Z.~Li, H.~Qian, C.~Yang, Z.~Wang, M.~Shang, V.~Kumar, S.~Tan, B.~Ray, P.~Bhatia, R.~Nallapati, M.~K. Ramanathan, D.~Roth, and B.~Xiang, ``{R}e{C}ode: Robustness evaluation of code generation models,'' in \emph{Proceedings of the 61st Annual Meeting of the Association for Computational Linguistics (Volume 1: Long Papers)}.\hskip 1em plus 0.5em minus 0.4em\relax Toronto, Canada: Association for Computational Linguistics, Jul. 2023, pp. 13\,818--13\,843. [Online]. Available: \url{https://aclanthology.org/2023.acl-long.773}
\BIBentrySTDinterwordspacing

\bibitem{mastropaolo2023robustness}
A.~Mastropaolo, L.~Pascarella, E.~Guglielmi, M.~Ciniselli, S.~Scalabrino, R.~Oliveto, and G.~Bavota, ``On the robustness of code generation techniques: An empirical study on github copilot,'' in \emph{ICSE}, 2023.

\bibitem{zhong2023study}
L.~Zhong and Z.~Wang, ``A study on robustness and reliability of large language model code generation,'' \emph{arXiv preprint arXiv:2308.10335}, 2023.

\bibitem{zhou2022adversarial}
Y.~Zhou, X.~Zhang, J.~Shen, T.~Han, T.~Chen, and H.~Gall, ``Adversarial robustness of deep code comment generation,'' \emph{ACM Transactions on Software Engineering and Methodology (TOSEM)}, vol.~31, no.~4, pp. 1--30, 2022.

\bibitem{zhuo2023robustness}
T.~Y. Zhuo, Z.~Li, Y.~Huang, Y.-F. Li, W.~Wang, G.~Haffari, and F.~Shiri, ``On robustness of prompt-based semantic parsing with large pre-trained language model: An empirical study on codex,'' \emph{arXiv preprint arXiv:2301.12868}, 2023.

\bibitem{shirafuji2023exploring}
A.~Shirafuji, Y.~Watanobe, T.~Ito, M.~Morishita, Y.~Nakamura, Y.~Oda, and J.~Suzuki, ``Exploring the robustness of large language models for solving programming problems,'' \emph{arXiv preprint arXiv:2306.14583}, 2023.

\bibitem{ling2016latent}
W.~Ling, P.~Blunsom, E.~Grefenstette, K.~M. Hermann, T.~Ko{\v{c}}isk{\`y}, F.~Wang, and A.~Senior, ``Latent predictor networks for code generation,'' in \emph{Proceedings of the 54th Annual Meeting of the Association for Computational Linguistics (Volume 1: Long Papers)}, 2016, pp. 599--609.

\bibitem{yin2017syntactic}
P.~Yin and G.~Neubig, ``A syntactic neural model for general-purpose code generation,'' in \emph{Proceedings of the 55th Annual Meeting of the Association for Computational Linguistics (Volume 1: Long Papers)}, 2017, pp. 440--450.

\bibitem{svyatkovskiy2020intellicode}
A.~Svyatkovskiy, S.~K. Deng, S.~Fu, and N.~Sundaresan, ``Intellicode compose: Code generation using transformer,'' in \emph{Proceedings of the 28th ACM Joint Meeting on European Software Engineering Conference and Symposium on the Foundations of Software Engineering}, 2020, pp. 1433--1443.

\bibitem{wang2021codet5}
Y.~Wang, W.~Wang, S.~Joty, and S.~C. Hoi, ``Codet5: Identifier-aware unified pre-trained encoder-decoder models for code understanding and generation,'' in \emph{Proceedings of the 2021 Conference on Empirical Methods in Natural Language Processing}, 2021, pp. 8696--8708.

\bibitem{le2022coderl}
H.~Le, Y.~Wang, A.~D. Gotmare, S.~Savarese, and S.~C.~H. Hoi, ``Coderl: Mastering code generation through pretrained models and deep reinforcement learning,'' \emph{Advances in Neural Information Processing Systems}, vol.~35, pp. 21\,314--21\,328, 2022.

\bibitem{codegen2}
E.~Nijkamp, H.~Hayashi, C.~Xiong, S.~Savarese, and Y.~Zhou, ``Codegen2: Lessons for training llms on programming and natural languages,'' \emph{arXiv preprint arXiv:2305.02309}, 2023.

\bibitem{codegeex}
Q.~Zheng, X.~Xia, X.~Zou, Y.~Dong, S.~Wang, Y.~Xue, Z.~Wang, L.~Shen, A.~Wang, Y.~Li \emph{et~al.}, ``Codegeex: A pre-trained model for code generation with multilingual evaluations on humaneval-x,'' \emph{arXiv preprint arXiv:2303.17568}, 2023.

\bibitem{codellama}
B.~Rozière, J.~Gehring, F.~Gloeckle, S.~Sootla, I.~Gat, X.~E. Tan, Y.~Adi, J.~Liu, T.~Remez, J.~Rapin, A.~Kozhevnikov, I.~Evtimov, J.~Bitton, M.~Bhatt, C.~C. Ferrer, A.~Grattafiori, W.~Xiong, A.~Défossez, J.~Copet, F.~Azhar, H.~Touvron, L.~Martin, N.~Usunier, T.~Scialom, and G.~Synnaeve, ``Code llama: Open foundation models for code,'' 2023.

\bibitem{sun2020treegen}
Z.~Sun, Q.~Zhu, Y.~Xiong, Y.~Sun, L.~Mou, and L.~Zhang, ``Treegen: A tree-based transformer architecture for code generation,'' in \emph{Proceedings of the AAAI Conference on Artificial Intelligence}, vol.~34, no.~05, 2020, pp. 8984--8991.

\bibitem{bleu}
K.~Papineni, S.~Roukos, T.~Ward, and W.-J. Zhu, ``Bleu: a method for automatic evaluation of machine translation,'' in \emph{Proceedings of the 40th annual meeting of the Association for Computational Linguistics}, 2002, pp. 311--318.

\bibitem{zhou2022docprompting}
S.~Zhou, U.~Alon, F.~F. Xu, Z.~Jiang, and G.~Neubig, ``Docprompting: Generating code by retrieving the docs,'' in \emph{The Eleventh International Conference on Learning Representations}, 2022.

\bibitem{li2023towards}
J.~Li, Y.~Zhao, Y.~Li, G.~Li, and Z.~Jin, ``Acecoder: Utilizing existing code to enhance code generation,'' 2023.

\bibitem{arxiv}
``arxiv.org e-print archive,'' \url{https://arxiv.org/}, 2023, last accessed October 2023.

\bibitem{replication_package}
``Link to our replication package,'' \url{https://github.com}, 2023, last accessed December 2023.

\bibitem{ding2023temporal}
Z.~Ding, Y.~Tang, Y.~Li, H.~Li, and W.~Shang, ``On the temporal relations between logging and code,'' in \emph{2023 IEEE/ACM 45th International Conference on Software Engineering (ICSE)}.\hskip 1em plus 0.5em minus 0.4em\relax IEEE, 2023, pp. 843--854.

\bibitem{deeplv}
Z.~Li, H.~Li, T.-H.~P. Chen, and W.~Shang, ``Deeplv: Suggesting log levels using ordinal based neural networks,'' in \emph{2021 IEEE/ACM 43rd International Conference on Software Engineering (ICSE)}.\hskip 1em plus 0.5em minus 0.4em\relax IEEE, 2021, pp. 1461--1472.

\bibitem{wheretolog_ASE20}
Z.~Li, T.~Chen, and W.~Shang, ``Where shall we log? studying and suggesting logging locations in code blocks,'' in \emph{35th {IEEE/ACM} International Conference on Automated Software Engineering, {ASE} 2020}, 2020, pp. 361--372.

\bibitem{kappa}
M.~L. McHugh, ``Interrater reliability: the kappa statistic,'' \emph{Biochemia Medica}, vol.~22, no.~3, pp. 276--282, 2012.

\bibitem{survey_guide_2008}
B.~A. Kitchenham and S.~L. Pfleeger, ``Personal opinion surveys,'' in \emph{Guide to advanced empirical software engineering}, 2008, pp. 63--92.

\bibitem{hu2022practitioners}
X.~Hu, X.~Xia, D.~Lo, Z.~Wan, Q.~Chen, and T.~Zimmermann, ``Practitioners' expectations on automated code comment generation,'' in \emph{Proceedings of the 44th International Conference on Software Engineering}, 2022, pp. 1693--1705.

\bibitem{naming_methods}
R.~Alsuhaibani, C.~Newman, M.~Decker, M.~Collard, and J.~Maletic, ``On the naming of methods: A survey of professional developers,'' in \emph{2021 IEEE/ACM 43rd International Conference on Software Engineering (ICSE)}, 2021, pp. 587--599.

\bibitem{microsoft_forms}
``Microsoft forms,'' \url{https://forms.office.com/}, 2023, last accessed February 2024.

\bibitem{mbpp_austin2021program}
J.~Austin, A.~Odena, M.~Nye, M.~Bosma, H.~Michalewski, D.~Dohan, E.~Jiang, C.~Cai, M.~Terry, Q.~Le \emph{et~al.}, ``Program synthesis with large language models,'' \emph{arXiv preprint arXiv:2108.07732}, 2021.

\bibitem{chen2024evaluating}
J.~Chen, Z.~Pan, X.~Hu, Z.~Li, G.~Li, and X.~Xia, ``Evaluating large language models with runtime behavior of program execution,'' \emph{arXiv preprint arXiv:2403.16437}, 2024.

\bibitem{huo2023autolog}
Y.~Huo, Y.~Li, Y.~Su, P.~He, Z.~Xie, and M.~R. Lyu, ``Autolog: A log sequence synthesis framework for anomaly detection,'' in \emph{2023 38th IEEE/ACM International Conference on Automated Software Engineering (ASE)}.\hskip 1em plus 0.5em minus 0.4em\relax IEEE, 2023, pp. 497--509.

\bibitem{li2023ase}
Z.~Li, A.~R. Chen, X.~Hu, X.~Xia, T.-H. Chen, and W.~Shang, ``Are they all good? studying practitioners' expectations on the readability of log messages,'' in \emph{2023 38th IEEE/ACM International Conference on Automated Software Engineering (ASE)}.\hskip 1em plus 0.5em minus 0.4em\relax IEEE, 2023, pp. 129--140.

\bibitem{li2023icse}
Z.~Li, C.~Luo, T.-H. Chen, W.~Shang, S.~He, Q.~Lin, and D.~Zhang, ``Did we miss something important? studying and exploring variable-aware log abstraction,'' in \emph{2023 IEEE/ACM 45th International Conference on Software Engineering (ICSE)}.\hskip 1em plus 0.5em minus 0.4em\relax IEEE, 2023, pp. 830--842.

\bibitem{dlfinder_TSE}
Z.~Li, T.-H. Chen, J.~Yang, and W.~Shang, ``Studying duplicate logging statements and their relationships with code clones,'' \emph{IEEE Transactions on Software Engineering}, pp. 2476--2494, 2021.

\bibitem{fried2022incoder}
D.~Fried, A.~Aghajanyan, J.~Lin, S.~Wang, E.~Wallace, F.~Shi, R.~Zhong, S.~Yih, L.~Zettlemoyer, and M.~Lewis, ``Incoder: A generative model for code infilling and synthesis,'' in \emph{The Eleventh International Conference on Learning Representations}, 2022.

\bibitem{luo2023wizardcoder}
Z.~Luo, C.~Xu, P.~Zhao, Q.~Sun, X.~Geng, W.~Hu, C.~Tao, J.~Ma, Q.~Lin, and D.~Jiang, ``Wizardcoder: Empowering code large language models with evol-instruct,'' 2023.

\bibitem{spacy}
``{spaCy}, {Explosion},'' \url{https://spacy.io/}, 2023, last accessed August 2023.

\bibitem{LemmInflect}
``{LemmInflect},{bjascob},'' \url{https://github.com/bjascob/LemmInflect}, 2023, last accessed August 2023.

\bibitem{fellbaum2010wordnet}
C.~Fellbaum, ``Wordnet,'' in \emph{Theory and applications of ontology: computer applications}.\hskip 1em plus 0.5em minus 0.4em\relax Springer, 2010, pp. 231--243.

\bibitem{wordnet_in_se}
X.~Chen, C.~Chen, D.~Zhang, and Z.~Xing, ``Sethesaurus: Wordnet in software engineering,'' \emph{IEEE Transactions on Software Engineering}, pp. 1960--1979, 2019.

\bibitem{liu2020senet}
Y.~Liu, J.~Lin, J.~Cleland-Huang, M.~Vierhauser, J.~Guo, and S.~Lohar, ``Senet: a semantic web for supporting automation of software engineering tasks,'' in \emph{2020 IEEE Seventh International Workshop on Artificial Intelligence for Requirements Engineering (AIRE)}, 2020, pp. 23--32.

\bibitem{bhatia2020clustering}
K.~Bhatia, S.~Mishra, and A.~Sharma, ``Clustering glossary terms extracted from large-sized software requirements using fasttext,'' in \emph{Proceedings of the 13th Innovations in Software Engineering Conference on Formerly known as India Software Engineering Conference}, 2020, pp. 1--11.

\bibitem{prithivida2021parrot}
P.~Damodaran, ``Parrot: Paraphrase generation for nlu.'' 2021.

\bibitem{openai_api}
``{OpenAI API},'' \url{https://platform.openai.com/docs/guides/gpt/chat-completions-api}, 2023, last accessed August 2023.

\bibitem{belinkov2018synthetic}
Y.~Belinkov and Y.~Bisk, ``Synthetic and natural noise both break neural machine translation,'' in \emph{International Conference on Learning Representations (ICLR)}, 2018.

\bibitem{dlfinder}
Z.~Li, T.~P. Chen, J.~Yang, and W.~Shang, ``{DLF}inder: characterizing and detecting duplicate logging code smells,'' in \emph{Proceedings of the 41st International Conference on Software Engineering, {ICSE} 2019}, 2019, pp. 152--163.

\end{thebibliography}

\end{document}